\documentclass[final]{IEEEtran}
\usepackage{cite}
\usepackage{threeparttable}
\usepackage{graphicx}
\usepackage{picinpar}
\usepackage{stfloats}
\usepackage{setspace}
\usepackage[all]{xy}
\usepackage{float}
\usepackage[cmex10]{amsmath}
\usepackage{amsmath,amsfonts,amssymb}
\usepackage{array}
\usepackage{mdwmath}
\usepackage{mdwtab}
\usepackage{eqparbox}
\usepackage{algorithmic}
\usepackage[ruled,lined,boxed]{algorithm2e}
\usepackage{subfigure}
\usepackage{threeparttable}
\usepackage{caption}
\usepackage{booktabs}
\usepackage{multirow}
\usepackage{etoolbox}
\usepackage{makecell}
\usepackage{amssymb}
\usepackage{mathrsfs}
\usepackage{color}
\definecolor{purple}{RGB}{160,32,240}
\definecolor{darkred}{RGB}{255,0,255}
\usepackage{stfloats}
\usepackage{bm}
\usepackage{flushend}
\usepackage[justification=centering]{caption}

\newcommand{\e}{\begin{equation}}
\newcommand{\ee}{\end{equation}}
\newcommand{\eqn}{\begin{eqnarray}}
\newcommand{\eeqn}{\end{eqnarray}}

\renewcommand{\raggedright}{\leftskip=0pt \rightskip=0pt plus 0cm}

\begin{document}

\title{Massive Access in Media Modulation Based Massive Machine-Type Communications}
\author{Li Qiao, Jun Zhang, Zhen Gao, Derrick Wing Kwan Ng,~\IEEEmembership{Fellow,~IEEE}, Marco Di Renzo,~\IEEEmembership{Fellow,~IEEE,} and Mohamed-Slim Alouini,~\IEEEmembership{Fellow,~IEEE}%

\vspace{-5.0mm}
\thanks{This paper was presented in part at the 2020 IEEE Wireless Communications and Networking Conference Workshops (WCNCW), 2020 \cite{WCNC}. The work was supported by NSFC under Grants 62071044 and 61827901, the BJNSF under Grant L182024. D. W. K. Ng is supported by funding from the UNSW Digital Grid Futures Institute, UNSW, Sydney, under a cross-disciplinary fund scheme and by the Australian Research Council's Discovery Project (DP210102169). The codes and some other materials about this work may be available at https://gaozhen16.github.io. (Corresponding Author: Jun Zhang and Zhen Gao)}
\thanks{L. Qiao, J. Zhang, and Z. Gao are with the School of Information and Electronics, Beijing Institute of Technology, Beijing 100081, China (e-mail: qiaoli@bit.edu.cn; buaazhangjun@vip.sina.com; gaozhen16@bit.edu.cn).}
\thanks{D. W. K. Ng is with the School of Electrical Engineering and Telecommunications, University of New South Wales, Sydney,
NSW 2025, Australia (email: w.k.ng@unsw.edu.au).}
\thanks{M. Di Renzo is with Universit\'e Paris-Saclay, CNRS, CentraleSup\'elec, Laboratoire des Signaux et Syst\`emes, 3 Rue Joliot-Curie, 91192 Gif-sur-Yvette, France (email: marco.di-renzo@universite-paris-saclay.fr).} %
\thanks{M.-S. Alouini is with the Computer, Electrical and Mathematical Science and Engineering Division, King Abdullah University of
Science and Technology (KAUST), Thuwal 23955-6900, Saudi Arabia (email: slim.alouini@kaust.edu.sa).}
}

\maketitle
\begin{abstract}
The massive machine-type communications (mMTC) paradigm based on media modulation in conjunction with massive multi-input multi-output base stations (BSs) is emerging as a viable solution to support the massive connectivity for the future Internet-of-Things, in which the inherent massive access at the BSs poses significant challenges for device activity and data detection (DADD). This paper considers the DADD problem for both uncoded and coded media modulation based mMTC with a slotted access frame structure, where the device activity remains unchanged within one frame. Specifically, due to the slotted access frame structure and the adopted media modulated symbols, the access signals exhibit a {\it doubly structured sparsity} in both the time domain and the modulation domain. Inspired by this, a doubly structured approximate message passing (DS-AMP) algorithm is proposed for reliable DADD in the uncoded case. Also, we derive the state evolution of the DS-AMP algorithm to theoretically characterize its performance. As for the coded case, we develop a bit-interleaved coded media modulation scheme and propose an iterative DS-AMP (IDS-AMP) algorithm based on successive inference cancellation (SIC), where the signal components associated with the detected active devices are successively subtracted to improve the data decoding performance. In addition, the channel estimation problem for media modulation based mMTC is discussed and an efficient data-aided channel state information (CSI) update strategy is developed to reduce the training overhead in block fading channels. Finally, simulation results and computational complexity analysis verify the superiority of the proposed DS-AMP algorithm over state-of-the-art algorithms in the uncoded case. Also, our results confirm that the proposed SIC-based IDS-AMP algorithm can enhance the data decoding performance in the coded case and verify the validity of the proposed data-aided CSI update strategy.
\end{abstract}

\begin{IEEEkeywords}
Massive access, media modulation, massive multi-input multi-output, massive machine-type communications.
\end{IEEEkeywords}

\IEEEpeerreviewmaketitle

\section{Introduction}

\IEEEPARstart{W}{ith} the advent of the Internet-of-Things (IoT), massive machine-type communications (mMTC) are expected to support various potential applications, including smart metering, surveillance and healthcare. In fact, mMTC are considered to be an indispensable component for future beyond 5G/6G networks \cite{overview1,overview2}. In stark contrast with conventional human-type communications (HTC) that are characterized by downlink transmissions with long packets and high data rates, mMTC are characterized by uplink transmissions with short packets from massively deployed machine type devices (MTDs) whose data traffic is sporadic \cite{overview3,overview4}. The unique traffic characteristics of mMTC indicate that the existing access solutions designed for HTC can not effectively support the massive access of MTDs, which necessitates low-latency and high-reliable massive access techniques with low-complexity signal processing algorithms.

\vspace{-4.5mm}
\subsection{Related Work}
Existing access solutions can be mainly divided into two categories: grant-based access approaches and grant-free access approaches \cite{overview1,overview2,overview3,overview4}. As for grant-based access solutions, allocating orthogonal radio resources to different active MTDs via some sophisticated scheduling algorithms is necessary before the uplink transmission. However, this costs extra signaling overhead for handling the granted signals. In practice, due to the limited orthogonal resources, it would be difficult to support a large number of MTDs by applying grant-based schemes, and the use of complicated resource scheduling algorithms can lead to a prohibitive signaling overhead and latency, leading even to congestion \cite{overview1,overview2,overview3,overview4,TcomChen}. As an alternative solution, grant-free access schemes have recently emerged and have drawn significant attention \cite{JSAC-Editor,BWang1,YangDU1,Profshim2,BWang2,YangDU2,TWOLEVEL,AMP-AUD,MBMMUD3,MBMMUD4}. In grant-free access systems, the MTDs can transmit in the uplink without waiting for permission, which significantly simplifies the uplink access procedure and hence reduces the access latency compared with grant-based access protocols.

Due to the limited radio resources but large number of potential MTDs, grant-free massive access schemes result in non-orthogonal transmissions, which makes the device activity and data detection (DADD) problem more difficult to handle. In particular, the DADD can be treated as an underdetermined linear problem. This implies that the application of classical linear least-squares (LS) and linear minimum mean square error (LMMSE) detectors to tackle the DADD problem would result in poor detection performance \cite{MUDCS1}. Fortunately, thanks to the sporadic traffic characteristics of MTDs in mMTC, the number of active MTDs is usually much smaller than the number of total MTDs in any given time interval, which motivates the application of compressive sensing (CS) techniques to design effective grant-free massive access schemes \cite{MUDCS2}.

Recently, several grant-free massive access schemes have been proposed. Specifically, by leveraging the block sparsity of the data traffic of mMTC in consecutive time slots, the authors of \cite{BWang1} and \cite{YangDU1} proposed a structured iterative support detection algorithm and a threshold-aided block sparsity adaptive subspace pursuit algorithm, respectively, to jointly detect the active MTDs and the corresponding transmitted data with improved performance. Furthermore, by exploiting the finite alphabet constraint of the transmit data as {\it a priori} information, a maximum {\it a posteriori} probability-based greedy algorithm was proposed in \cite{Profshim2} to further improve the performance. In addition, in contrast to the greedy algorithms mentioned above, the authors of \cite{AMP-AUD} proposed an approximate message passing (AMP) algorithm based detector \cite{AMP,AMPmeng,KeMaLongTSP}, whereby the finite alphabet constraint of the transmit symbols was considered in the {\it a priori} probability and the expectation maximization (EM) algorithm was adopted to detect the active MTDs \cite{EM}. It is worth noting that the aforementioned works, i.e., \cite{BWang1,YangDU1,Profshim2,AMP-AUD}, considered a slotted access frame structure, where the MTDs remain (in)active over an entire data frame (several successive time slots). Yet, the authors of \cite{BWang2} and \cite{YangDU2} considered MTDs with dynamic device activity, where the activity of each MTD varies in several continuous time slots. Specifically, a dynamic CS-based detector and a prior-information-aided adaptive CS-based detector were proposed in \cite{BWang2} and \cite{YangDU2}, respectively. In both cases, the detection accuracy was improved by exploiting previously detected results. However, only single-antenna MTDs and single-antenna base stations (BSs) were considered in \cite{BWang1,YangDU1,Profshim2,BWang2,YangDU2,AMP-AUD}. 

To unleash the potential of massive access in mMTC, spatial modulation based on multiple transmit antennas at the MTDs and massive multi-input multi-output (mMIMO) BSs were considered in \cite{TWOLEVEL ,Gao}, and \cite{AMP-SM}. In particular, spatial modulation is a low-complexity and energy-efficient multiple-antenna scheme that utilizes a single or fewer radio frequency (RF) chains than the number of antenna elements \cite{SM1,SM2,SM3,SM4}. By encoding part of the information bits onto the activated antenna elements, spatial modulation is capable of enhancing the data rate at a low cost and low power consumption \cite{SM2}. The application and suitability of spatial modulation to the IoT is discussed and proved experimentally in \cite{SM6} and \cite{SM7}, respectively. In addition, the use of mMIMO at the BSs can significantly reduce the detection error probability of the active MTDs, thus improving the reliability of massive access \cite{YuWei}. Inspired by this, the authors of \cite{TWOLEVEL} proposed a two-level sparse structure CS (TLSSCS) algorithm for grant-free access, and the authors of \cite{Gao} and \cite{AMP-SM} considered a grant-based access scheme, where a group subspace pursuit algorithm and a structured AMP algorithm were proposed for detecting the signals encoded by using spatial modulation, respectively. By exploiting the structured sparsity of the spatial modulated symbols, the algorithms proposed in \cite{TWOLEVEL ,Gao}, and \cite{AMP-SM} achieve better data detection performance than the algorithms designed for single-antenna devices in \cite{BWang1,YangDU1,Profshim2,BWang2,YangDU2,AMP-AUD}. However, the spatial modulation transmission scheme adopted in \cite{TWOLEVEL ,Gao}, and \cite{AMP-SM} is the simplest one, which doubles the number of antenna elements for each extra spatial modulated data bit \cite{SM1}.

Fortunately, during the last years many improved transmission schemes have been proposed in order to enhance the spectral efficiency of spatial modulation without compromising its low-complexity and energy-efficiency. A recent overview and comparison of the most popular solutions, which includes generalized spatial modulation, media modulation, and, more recently, metasurface-based modulation, is available in \cite{OverviewMBM}. In this paper, we consider media modulation introduced in \cite{MBM1}. In particular, media modulation employs a single RF chain, a single radiating element, and several low-cost RF mirrors \cite{MBM2,MBM3,MBM4,MBM5}. The information bits are encoded into the active/inactive (or ON/OFF) status of the RF mirrors, which determines the resulting radiation pattern of the entire structure \cite{MBM2,MBM3}. In contrast to spatial modulation, the number of spatial bits encoded in media modulation is larger and depends on the number of distinguishable radiation patterns that can be realized \cite{MBM6}. Due to the promising advantages of media modulation, it was recently adopted in \cite{MBMMUD1,MBMMUD2,MBMMUD3,MBMMUD4} for application to the uplink transmission of MTDs with mMIMO BSs. Specifically, as for grant-based access schemes, the authors of \cite{MBMMUD1} and \cite{MBMMUD2} proposed an iterative interference cancellation detector and a message passing algorithm based detector, respectively. \textcolor{black}{As for grant-free access schemes with a slotted access frame structure, a structured orthogonal matching pursuit (StrOMP) algorithm for activity detection and a successive interference cancellation (SIC)-based structured subspace pursuit (SSP) algorithm for data detection were proposed in \cite{MBMMUD3}. In addition, a prior-information aided media modulation subspace matching pursuit (PIA-MSMP) algorithm was proposed in \cite{MBMMUD4} for DADD.} Additionally, the authors of \cite{MBMMUD4} proposed a prior-information aided adaptive media modulation subspace matching pursuit algorithm to accommodate the dynamic device activity. Although the structured sparsity of media modulated symbols was exploited for better performance in \cite{MBMMUD3} and \cite{MBMMUD4}, the proposed greedy algorithms failed to fully exploit the finite alphabet constraint of the transmit symbols. Hence, there are plenty of opportunities for further improving the design and optimization of media modulation for application to grant-free massive access.
It is important to note, in addition, that the grant-free massive access schemes in the aforementioned research works, i.e.,\cite{BWang1,YangDU1,Profshim2,BWang2,YangDU2,AMP-AUD,TWOLEVEL,Gao,AMP-SM,MBMMUD1,MBMMUD2,MBMMUD3,MBMMUD4}, are developed for an uncoded scenario, while their extension to coded systems needs further research.

\vspace{-3.5mm}
\subsection{Our Contributions}
This paper considers both uncoded and coded media modulation based mMTC, by assuming a slotted access frame structure. It is assumed that the activity of the MTDs remains unchanged in each frame, which yields a sparse transmission frame. To design a reliable DADD scheme, we introduce a doubly structured AMP (DS-AMP) algorithm for uncoded transmission. The state evolution (SE) of the proposed DS-AMP algorithm is derived in order to characterize its performance. As for the coded transmission, we introduce a bit-interleaved coded media modulation (BICMM) scheme and propose an SIC-based iterative DS-AMP (IDS-AMP) scheme for improving the decoding performance. In addition, the channel estimation (CE) problem for media modulation based mMTC is discussed and an effective channel state information (CSI) update strategy is developed to reduce the training overhead in block fading channels. With the aid of numerical simulations and computational complexity analysis, we verify the superiority of the proposed DS-AMP algorithm and the SIC-based IDS-AMP algorithm for uncoded and coded transmission, respectively, with respect to state-of-the-art benchmark schemes. Moreover, the effectiveness of the CSI update strategy is verified. The main contributions of this paper can be summarized as follows:

\begin{itemize}
\item[$\bullet$] {\bf DS-AMP algorithm for uncoded media modulation based mMTC:} By utilizing the structured sparsity of media modulated symbols in the modulation domain and the discrete distribution of quadrature amplitude modulation (QAM) alphabets, the proposed DS-AMP algorithm can reliably perform DADD. Furthermore, by leveraging the structured sparsity of the slotted access frame structure in the time domain, the active/inactive status of the MTDs and the noise variance can be adaptively learned via the EM algorithm with enhanced accuracy. Besides, we derive the theoretical SE, which closely matches the simulated results for the DS-AMP algorithm.

\item[$\bullet$] {\bf BICMM designed for coded media modulation based mMTC:} We integrate the bit-interleaved coded modulation (BICM) \cite{BICM,BICSM} into media modulation and develop a BICMM scheme, which can effectively mitigate the error bursts in fading channels and improve the data decoding performance. Particularly, we employ a bit-wise interleaver between the channel encoder and the media modulation module, where the bits of the media modulated symbols and the QAM symbols are collectively processed by channel coding and interleaving.

\item[$\bullet$] {\bf SIC-based IDS-AMP scheme for coded media modulation based mMTC:} We design a dedicated data packet structure for MTDs and develop an IDS-AMP detector at the BSs. The developed data packet includes a signature sequence part that is known by the transceiver and a payload part for data transmission. Particularly, we consider the decoding quality of the signature sequence as a metric to determine the SIC order. Different from the DS-AMP algorithm that yields hard-decision estimates, the developed IDS-AMP detector yields soft log-likelihood ratio (LLR) for both the media modulated and the QAM symbol bits, which are fed to the channel decoding module for improving the decoding accuracy. Furthermore, the SIC operation in the IDS-AMP detector successively subtracts the signal components associated with the well decoded active devices to improve the data decoding performance.
\end{itemize}

\textit {Notation}:  Matrices and vectors are denoted by symbols in boldface. Operators $\left(\cdot\right)^T$, $\left(\cdot\right)^*$, and $\left(\cdot\right)^{-1}$ represent transpose, conjugate, and inverse, respectively.
For a matrix ${\bf A}$ and an ordered set $\Omega$, ${\bf{A}}_{[m,n]}$ denote the $m$-th row and $n$-th column element of ${\bf{A}}$, ${\bf{A}}_{[\Omega,:]}$ (${\bf{A}}_{[:,\Omega]}$) is the sub-matrix containing the rows (columns) of ${\bf{A}}$ indexed by $\Omega$, and ${\bf{A}}_{[\Omega,m:n]}$ is the sub-matrix containing from the $m$-th to the $n$-th columns of ${\bf{A}}_{[\Omega,:]}$. For a vector ${\bf x}$ and an ordered set $\Omega$, ${\left\| {\bf x} \right\|_p}$, $[{\bf x}]_{m}$, and $[{\bf x}]_{\Omega}$ denote the ${l_p}$ norm, $m$-th element of ${\bf x}$, and the elements of ${\bf x}$ indexed by $\Omega$, respectively. $\Theta({\bf x},n)$ is the operator that selects the indices of the top $n$ largest elements of ${\bf x}$. For an ordered set $\Gamma$ and its subset $\Omega$, $|\Gamma|_c$, $\Gamma|_i$, and $\Gamma\setminus \Omega$ denote the cardinality of $\Gamma$, the $i$-th element of $\Gamma$, and the complement of subset $\Omega$ in $\Gamma$, respectively. For the binary codewords ${\bf c}_1$ and ${\bf c}_2$, we define their Hamming distance $D({\bf c}_1,{\bf c}_2)$ as the number of elements in which they differ. $[K]$ is the set $\{1,2,...,K\}$ and ${\bf 0}_{m\times n}$ denotes a zero matrix with size $m\times n$. If $h$ is the joint distribution of variable ${\bf x}=[x_1,x_2,x_3]^T$, then the marginal distribution of $x_2$ is denoted by $\sum\nolimits_{\sim\{x_2\}}h({\bf x})=\sum\nolimits_{x_1\in\mathbb{A}}\sum\nolimits_{x_3\in\mathbb{A}}h({\bf x})$, where $x_i$, $\forall i\in\{1,2,3\}$, belongs to a finite domain $\mathbb{A}$. Finally, $\mathcal{CN}(x; \mu, \nu)$ denotes the complex Gaussian distribution of a random variable $x$ with mean $\mu$ and variance $\nu$.

\begin{figure*}[t]
\vspace{-6mm}
\captionsetup{font={footnotesize}, singlelinecheck = off, justification = raggedright,name={Fig.},labelsep=period}
\centering
\begin{minipage}[c]{0.49\textwidth}
\centerline{\includegraphics[scale=.29]{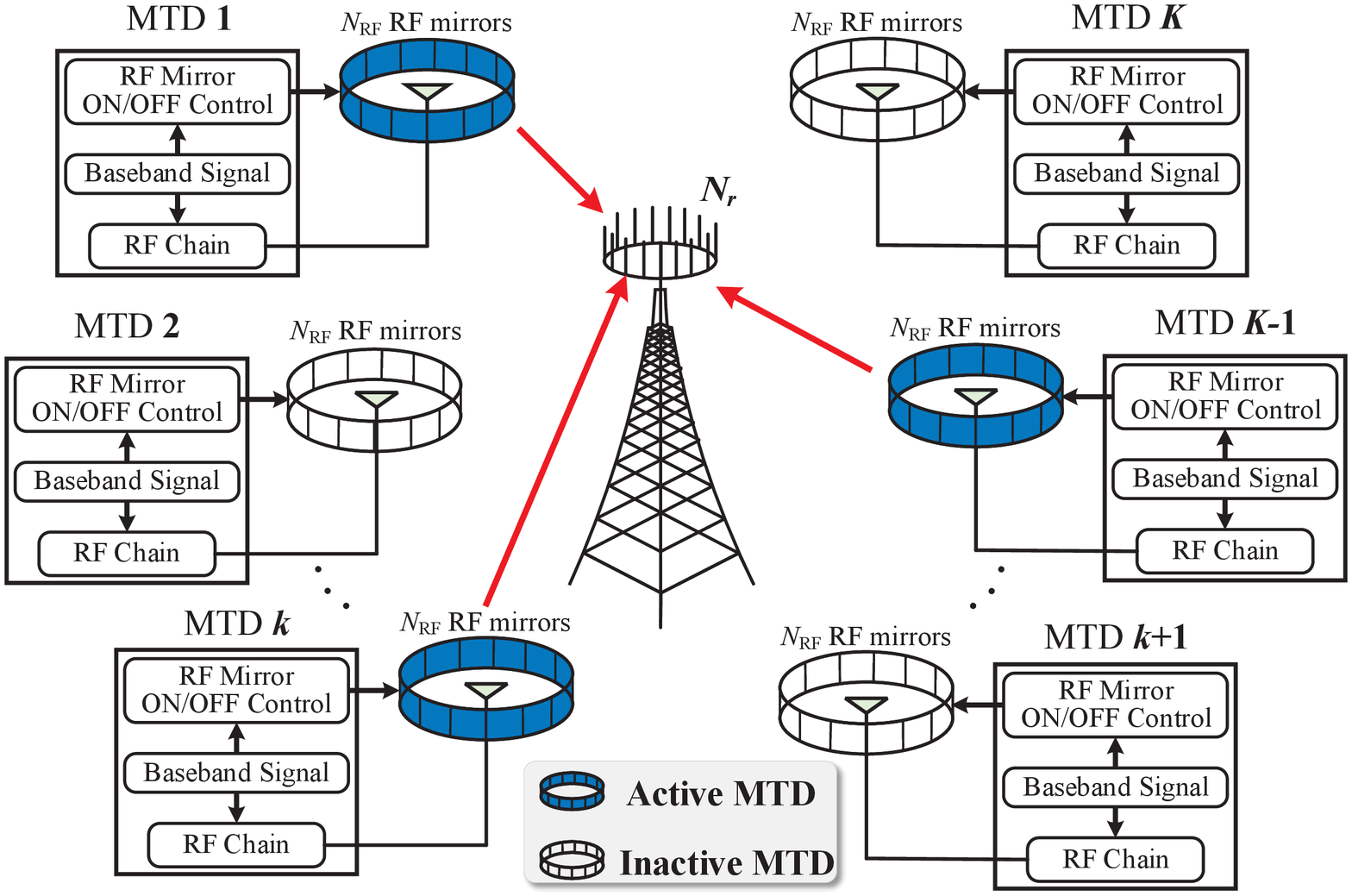}}
\caption{Media modulation based mMTC scheme: The MTDs adopt media modulation to access the mMIMO BSs.}
\label{Fig:1}
\end{minipage}
\hfill
\begin{minipage}[c]{0.49\textwidth}
\centering%
\centerline{\includegraphics[scale=.4]{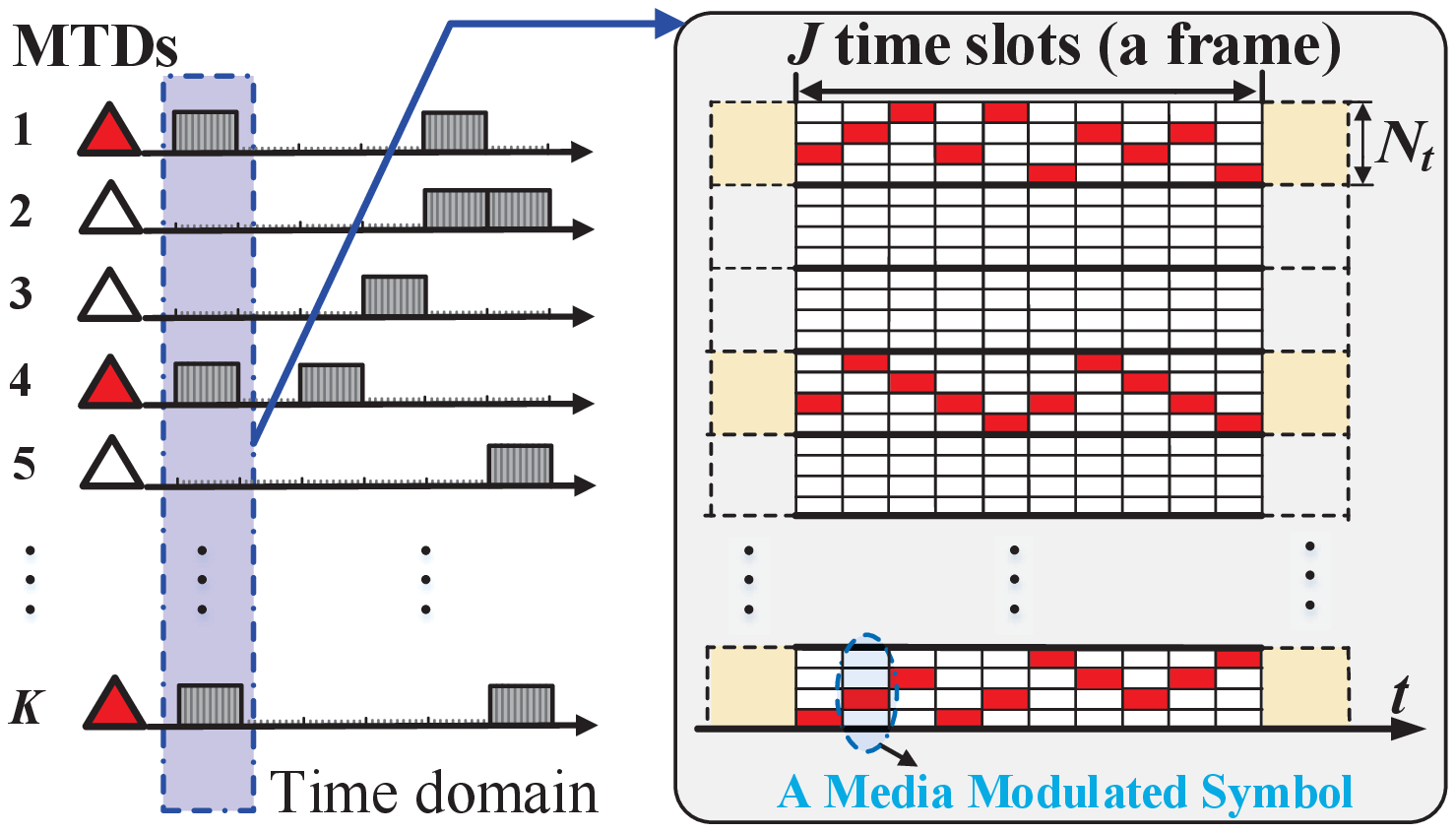}}
\caption{Media modulation based mMTC with a slotted access frame structure, where the invariant active/inactive status of the MTDs within a frame forms a {\it structured sparsity in the time domain}, and the media modulated symbols possess a {\it structured sparsity in the modulation domain}.}
\label{Fig:2}
\end{minipage}
\vspace{-6mm}
\end{figure*}
%
\vspace{-1.5mm}
\section{System Model}

In this section, we first introduce the media modulation based mMTC scheme and then focus on the massive access at the BS for DADD.

\vspace{-3mm}
\subsection{Media Modulation Based mMTC Scheme}

As illustrated in Fig. \ref{Fig:1}, we consider $K$ MTDs that employ media modulation for enhancing the throughput and the BS employs mMIMO with $N_r\gg 1$ antenna elements for realizing a reliable massive access. Only $K_a$ out of $K$ ($K\!\gg\!K_a$) MTDs are active simultaneously. Specifically, each MTD is equipped with one RF chain, one transmit antenna, and $N_{\rm RF}$ low-cost RF mirrors, where each RF mirror has a controllable binary ON/OFF status. Each device has $N_t=2^{N_{\rm RF}}$ different kinds of mirror activation patterns (MAPs), i.e., $N_t$ different channel realizations, which can be used to encode ${\rm log_2}{N_t}=N_{\rm RF}$ extra information bits. As for the uplink access, therefore, the throughput is $\eta ={\rm log_2} M+N_{\rm RF}$ bit per channel use (bpcu), which consists of ${\rm log_2} M$ bits conveyed by, e.g., an $M$-ary QAM ($M$-QAM) scheme and $N_{\rm RF}$ bits conveyed by media modulation. Furthermore, the use of an mMIMO offers improved detection performance for massive access \cite{Gao,YuWei}.

\vspace{-3mm}
\subsection{Massive Access in Media Modulation Based mMTC}

As shown in Fig. \ref{Fig:2}, we consider a media modulation based mMTC system with a slotted access frame structure, where the active/inactive status of the $K$ MTDs is invariable in a frame, i.e., $J$ successive time slots. We refer to this property as the {\it structured sparsity in the time domain}. For convenience, we focus our attention on the massive access problem in any given frame without loss of generality. Specifically, the received signal at the BS in the $j$-th, $\forall j\in [J]$, time slot, which is denoted by ${\bf{y}}_{j}\in\mathbb{C}^{N_r \times 1}$, can be written as
\begin{equation}\label{eq:system}
\begin{split}
{{\bf{y}}_{j}}&=\sum\nolimits_{k = 1}^K a_ks_{k,j}{{\bf{H}}_k{{\bf{d}}_{k,j}} + } {{\bf{w}}_{j}}\\
&=\sum\nolimits_{k = 1}^K {{\bf{H}}_k{{\bf{x}}_{k,j}} + } {{\bf{w}}_{j}}
={\bf{{{H}}}}{{\bf{{\tilde{x}}}}_{j}} +  {{\bf{w}}_{j}},
\end{split}
\end{equation}
where the binary activity indicator $a_k\in\{0,1\}$ is one (zero) if the $k$-th MTD is active (inactive), $s_{k,j}$ of the $k$-th MTD in the $j$-th time slot is selected from the $M$-QAM set $\mathbb{S}$, ${\bf d}_{k,j}\in\mathbb{C}^{N_t \times 1}$ and ${\bf{x}}_{k,j}=a_k{s_{k,j}}{\bf d}_{k,j}\in\mathbb{C}^{N_t \times 1}$ are the media modulated symbol and the effective uplink transmitted symbols of the $k$-th MTD in the $j$-th time slot, respectively, ${\bf{H}}_k\in\mathbb{C}^{N_r\times N_t}$ is the multiple-input multiple-output (MIMO) channel matrix corresponding to the $k$-th MTD, ${\bf{w}}^{j} \in \mathbb{C}^{N_r \times 1}$ is the Gaussian noise whose elements follow an independent and identically distributed (i.i.d.) complex Gaussian distribution $\mathcal{CN}([{\bf{w}}^{j}]_n;0,\sigma_w^2)$, $\forall n\in[N_r]$, ${{\bf{ H}}} = [{\bf H} _1, {\bf H} _2,...,{\bf H} _K] \in \mathbb{C}^{N_r \times (K N_t)}$ and ${{\bf{\tilde x}}_{j} } = [({\bf x}_{1,j})^{ T},({\bf x}_{2,j})^{ T},...,({\bf x}_{K,j})^{ T}]^{ T} \in \mathbb{C}^{(K N_t) \times 1}$ are the aggregated channel matrix and uplink access signal of the $j$-th time slot, respectively.

According to the media modulation transmission scheme, only one entry of the media modulated symbol ${\bf d}_{k,j}$, $\forall j\in[J]$ and $\forall k\in[K]$, is one and the others are zeros, i.e.,
\begin{equation}\label{eq:d}
\begin{array}{l}
{\rm supp}\left\{{\bf d}_{k,j}\right\}\in [N_t],~~\left\|{{{\bf d}_{k,j}}}\right\|_0=1,~~\left\|{{{\bf d}_{k,j}}}\right\|_2=1,
\end{array}
\end{equation}
where ${\rm supp}\{{\bf d}_{k,j}\}$ denotes the support set of ${\bf d}_{k,j}$. We refer to this property as the {\it structured sparsity in the modulation domain}.
Additionally, we consider the commonly adopted Gauss-Markov block fading channel model \cite{BlockFading-JSTSP,BlockFading-ACCESS}, and the details of the CE problem will be discussed in Section V.

\section{Proposed Solution for Uncoded Media Modulation Based mMTC}

In this section, we present the proposed DS-AMP algorithm for DADD in uncoded media modulation based mMTC. First, we exploit the {\it doubly structured sparsity} in media modulation based mMTC and formulate an optimization problem for massive access. Then, by exploiting the {\it doubly structured sparsity} as {\it a priori} information, we propose a DS-AMP algorithm for effective DADD. Subsequently, the SE of the DS-AMP algorithm is derived to theoretically predict the performance. Finally, the computational complexity of the DS-AMP algorithm is analyzed.

\vspace{-2mm}
\subsection{Preliminaries}
Although the total number of MTDs in the IoT is generally massive, their sporadic traffic behavior results in a sparse activity, i.e., at any given time slot, only a minority of MTDs are active \cite{overview1,overview2,overview3,overview4}. Hence, we introduce an activity indicator vector ${\bf a}=[a_1,a_2,...,a_K]^T\in\mathbb{C}^{K \times 1}$, which is sparse as the number of active MTDs $K_a=\left\|{{\bf a}}\right\|_0\ll K$. Furthermore, we collectively refer to the {\it structured sparsity in the time domain} due to the slotted access frame structure and the {\it structured sparsity in the modulation domain} shown in (\ref{eq:d}) as the {\it doubly structured sparsity}.

To exploit the {\it structured sparsity in the time domain}, we rewrite the received signals of $J$ successive time slots in a compact matrix form as
\begin{equation}\label{eq:systemModel}
\begin{array}{l}
\bf{Y}=\bf{ H}\bf{X}+\bf{W},
\end{array}
\end{equation}
where we have ${\bf{Y}}\!=\![{\bf {y}}_{1}, {\bf {y}}_{2}, ..., {\bf {y}}_{J}]\in\mathbb{C}^{N_r \times J}$, ${{\bf{H}}}\in\mathbb{C}^{N_r \times (K N_t)}$, ${\bf{X}}\!=\![{{ {{\bf\tilde{x}}_{1}}}}, {{ {{\bf\tilde{x}}_{2}}}}, ..., {{{\bf \tilde {x}}_{J}}}]\in\mathbb{C}^{(K N_t) \times J}$,
and ${\bf{W}}=[{{ {{\bf{w}}_{1}}}}, {{ {{\bf{w}}_{2}}}}, ..., {{{\bf {w}}_{J}}}]\in\mathbb{C}^{N_r \times J}$.
Hence the massive access problem can be formulated as the following optimization problem
\begin{align}\label{eq:OPTproblem}
&\min\limits_{\bf X} \left\|{ {\bf Y}-{\bf HX}}\right\|_F^2=\min\limits_{\{{\bf \tilde x}_{j}\}_{j=1}^{J}}\sum\limits_{j=1}^J\left\|{ {\bf y}_j-{\bf  H}{\bf \tilde x}_j}\right\|_2^2\nonumber \\
&=\min\limits_{\{a_k,{\bf d}_{k,j},s_{k,j}\}_{j=1,k=1}^{J,K}} \,\, \sum\limits_{j=1}^J\left\|{ {\bf y}_j-\sum\limits_{k = 1}^K a_k s_{k,j} {\bf H}_k{\bf d}_{k,j}}\right\|_2^2\nonumber\\
&{\rm s.t.}~(2),~\left\|{\bf a}\right\|_0\ll K,~{\rm and}~s_{k,j}\in\mathbb{S}, k\in[K],j\in[J].
\end{align}

\vspace{-4mm}
\subsection{Proposed DS-AMP Algorithm for DADD}
\subsubsection{Problem Formulation Based on Factor Graph}
The optimization problem in (\ref{eq:OPTproblem}) minimizes the mean square error between ${\bf Y}$ and ${\bf HX}$, which is equivalent to estimating the {\it a posteriori} mean of the uplink access signal ${\bf X}$ \cite{Kay}{\footnote{\color{black}Due to space limitations, interested readers are referred to Eqs. (10.2)-(10.5) of Chapter 10 in\cite{Kay} for the detailed derivations.}}. In (\ref{eq:OPTproblem}), the {\it a posteriori} mean of $\left[{{\bf x}_{k,j}}\right]_i$, $\forall k\in[K]$, $\forall j\in[J]$, $\forall i\in[N_t]$, can be expressed as

\begin{equation}\label{eq:PosMeanGeneral}
\begin{array}{l}
\left[{\widehat{\bf x}_{k,j}}\right]_i=\sum\limits_{\left[{{\bf x}_{k,j}}\right]_i\in \mathbb{\overline S}}\left[{{\bf x}_{k,j}}\right]_i p\left({\left[{{\bf x}_{k,j}}\right]_i|{\bf y}_j}\right),
\end{array}
\end{equation}
where $\mathbb{\overline S}=\left\{{\mathbb{S},0}\right\}$, $p\left({\left[{{\bf x}_{k,j}}\right]_i|{\bf y}_j}\right)$ is the marginal distribution of $p\left({\tilde{\bf x}_j|{\bf y}_j}\right)$ that can be expressed as
\begin{equation}\label{eq:MarginalGenaral}
\begin{array}{l}
p\left({\left[{{\bf x}_{k,j}}\right]_i|{\bf y}_j}\right)=\sum\limits_{\sim\left\{{\left[{{\bf x}_{k,j}}\right]_i}\right\}} p\left({\tilde{\bf x}_j|{\bf y}_j}\right).
\end{array}
\end{equation}

Based on Bayes' theorem, the joint posterior distribution $p\left({\tilde{\bf x}_j|{\bf y}_j}\right)$ can be expressed as
\begin{align}\label{eq:PosDistriGeneral}
p\left({\tilde{\bf x}_j|{\bf y}_j;\sigma_w^2, {\bf a}}\right)&=\dfrac{p\left({{\bf y}_j|\tilde{\bf x}_j;\sigma_w^2}\right)p\left({\tilde{\bf x}_j;{\bf a}}\right)}{p\left({{\bf y}_j}\right)}\nonumber\\
&=\dfrac{1}{p\left({{\bf y}_j}\right)} \prod\limits_{n=1}^{N_r}p\left({\left[{{\bf y}_j}\right]_n|\tilde{\bf x}_j;\sigma_w^2}\right) \prod\limits_{k=1}^{K}p\left({{\bf x}_{k,j}; a_k}\right),
\end{align}
where the likelihood function can be expressed as
\begin{equation}\label{eq:LikelihoodGeneral}
\begin{array}{l}
p\left({\left[{{\bf y}_j}\right]_n|\tilde{\bf x}_j;\sigma_w^2}\right)\!\!=\!\!\dfrac{1}{\pi \sigma_w^2}{\rm exp}\left({-\dfrac{1}{\sigma_w^2}\left| {\left[{{\bf y}_j}\right]_n\!\!-\!\!\sum\limits_{k=1}^{K}\left[{{\bf H}_k{\bf x}_{k,j}}\right]_n} \right|^2}\right).
\end{array}
\end{equation}

According to the sparse device activity, the {\it structured sparsity in the modulation domain} in (\ref{eq:d}), and the discrete distribution of the QAM alphabet, the {\it a priori} distribution $p\left({{\bf x}_{k,j}; a_k}\right)$ in (\ref{eq:PosDistriGeneral}) is formulated as
\begin{align}\label{eq:PriorGeneral}
&p\left({{\bf x}_{k,j}; a_k}\right)=(1-a_k)\prod\limits_{i=1}^{N_t}\delta\left({\left[{{\bf x}_{k,j}}\right]_i}\right)+\nonumber\\
&a_k\left\{{\dfrac{1}{N_t}\sum\limits_{i=1}^{N_t}\left[{\dfrac{1}{M}\sum\limits_{s\in\mathbb{S}}\delta\left({\left[{{\bf x}_{k,j}}\right]_i-s}\right)\prod\limits_{g\in [N_t],g\neq i}\delta\left({\left[{{\bf x}_{k,j}}\right]_g}\right)}\right]}\right\},
\end{align}
where $M=|\mathbb{S}|_c$ and $\delta\left({\cdot}\right)$ is the Dirac delta function.

\begin{figure}[t]
     \centering
     \includegraphics[width=7.5cm, keepaspectratio]
     {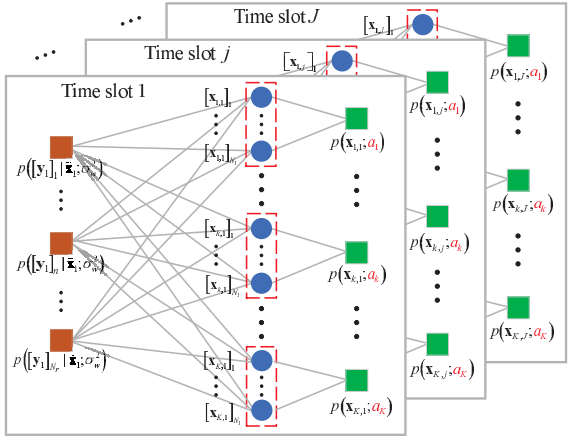}
     \vspace*{-1.3mm}
     \captionsetup{font={footnotesize}, singlelinecheck = off, justification = raggedright,name={Fig.},labelsep=period}
     \caption{Factor graph of the joint posterior distribution $p\left({\tilde{\bf x}_j|{\bf y}_j;\sigma_w^2, {\bf a}}\right)$, where the activity of the $k$-th MTD, $\forall k\in[K]$, remains unchanged in $J$ successive time slots and is denoted by $a_k$. The circles represent the variable nodes and the squares represent the factor nodes.}
     \label{fig:FactorGraph}
     \vspace*{-4mm}
\end{figure}

A factor graph representation of the joint posterior distribution $p\left({\tilde{\bf x}_j|{\bf y}_j;\sigma_w^2, {\bf a}}\right)$ in (\ref{eq:PosDistriGeneral}) is illustrated in Fig. \ref{fig:FactorGraph}. However, calculating the marginal distribution $p\left({\left[{{\bf x}_{k,j}}\right]_i|{\bf y}_j}\right)$, $\forall k\in[K]$, $\forall j\in[J]$, and $\forall i\in[N_t]$, from the joint posterior distribution $p\left({\tilde{\bf x}_j|{\bf y}_j;\sigma_w^2, {\bf a}}\right)$ is extremely complicated in massive access, due to the exceedingly large value of $K$. Fortunately, the AMP algorithm can provide an effective approximation of the marginal distributions at a low complexity, yet achieving near MMSE performance. In particular, by applying the AMP algorithm on the factor graph illustrated in Fig. \ref{fig:FactorGraph}, we propose a DS-AMP algorithm for handling the massive access problem. Specifically, we apply the DS-AMP algorithm to calculate the {\it a posteriori} mean of the media modulated signals ${\bf x}_{k,j}$ ($k\in[K]$, $j\in[J]$), and resort to the EM algorithm for estimating the activity indicators $a_k$ and the variance $\sigma_w^2$ of the complex Gaussian noise.

\subsubsection{Update Rules of the DS-AMP Algorithm}
As discussed in \cite{AMP,AMPmeng,KeMaLongTSP}, in the large system regime\footnote{Although developed in the large system regime, the AMP algorithm performs well even for medium size problems, such as massive access with hundreds or thousands of MTDs\cite{AMP,KeMaLongTSP}.} (i.e., $K\rightarrow\infty$, $\lambda=\frac{K_a}{K}$ and $\kappa=\frac{N_r}{K}$ are fixed), the AMP algorithm decouples, in the asymptotic regime, the matrix estimation problem of (\ref{eq:systemModel}) into $KJN_t$ uncoupled scalar problems
\begin{equation}\label{eq:systemdecoupled}
\begin{array}{l}
{\bf{Y}=\bf{ H}\bf{X}+{\bf{W}}}~\rightarrow~r_{l,j}=\left[{{\bf x}_{k,j}}\right]_i+\widehat{w}_{l,j},~\forall i,j,k,
\end{array}
\end{equation}
where $l=(k-1)N_t+i$, $r_{l,j}$ is the mean of $\left[{{\bf x}_{k,j}}\right]_i$ estimated by the AMP algorithm, $\widehat{w}_{l,j}\sim{\cal C}{\cal N}\left({{\widehat{w}_{l,j};0,\phi_{l,j}}}\right)$ is the equivalent noise, and $\phi_{l,j}$ is its variance.

In addition, the joint posterior distribution (\ref{eq:PosDistriGeneral}) can be approximated as
\begin{align}
\label{eq:PosDistriApproxi}p\left({\tilde{\bf x}_j|{\bf y}_j;\sigma_w^2, {\bf a}}\right)&\approx q\left({\tilde{\bf x}_j|{\bf y}_j;\sigma_w^2, {\bf a}}\right)\nonumber\\
&=\prod\limits_{k=1}^{K}\prod\limits_{i=1}^{N_t}q\left({\left[{{\bf x}_{k,j}}\right]_i|r_{l,j}, \phi_{l,j};\sigma_w^2, a_k}\right).
\end{align}

Based on Bayes' theorem, the approximated marginal {\it a posteriori} distribution, denoted as $q\left({\left[{{\bf x}_{k,j}}\right]_i|r_{l,j}, \phi_{l,j};\sigma_w^2, a_k}\right)$, $\forall i,j,k$, is given as follows
\begin{equation}\label{eq:PosMarginalAMP}
\begin{array}{l}
q\left({\left[{{\bf x}_{k,j}}\right]_i|r_{l,j}, \phi_{l,j};\sigma_w^2, a_k}\right)\\
~~~~~~~~=\dfrac{1}{q\left({r_{l,j};\sigma_w^2, a_k}\right)}q\left({r_{l,j}|\left[{{\bf x}_{k,j}}\right]_i;\sigma_w^2}\right)~p\left({\left[{{\bf x}_{k,j}}\right]_i;a_k}\right),
\end{array}
\end{equation}
where
\begin{align}
\label{eq:LikelihoodAMP} q\left({r_{l,j}|\left[{{\bf x}_{k,j}}\right]_i}\right)&=\dfrac{1}{\pi\phi_{l,j}}{\rm exp}\left({-\dfrac{1}{\phi_{l,j}}\left|{r_{l,j}-\left[{{\bf x}_{k,j}}\right]_i}\right|^2}\right),\\
\label{eq:PriorMarginal} p\left({\left[{{\bf x}_{k,j}}\right]_i;a_k}\right)&=\sum\nolimits_{\sim\left\{{\left[{{\bf x}_{k,j}}\right]_i}\right\}} p\left({{\bf x}_{k,j};a_k}\right)\nonumber\\
&=\left(\!\!{1\!-\!\frac{a_k}{N_t}}\!\!\right)\!\!\delta\left({\left[{{\bf x}_{k,j}}\right]_i}\right)\!\!+\!\!\frac{a_k}{N_tM}\!\!\sum\limits_{s\in\mathbb{S}}\delta\left(\!{\left[{{\bf x}_{k,j}}\right]_i\!-\!s}\!\right),\\
\label{eq:NormulizeItem} q\left({r_{l,j};\sigma_w^2, a_k}\right)&=\!\!\sum\limits_{\left[{{\bf x}_{k,j}}\right]_i\in\mathbb{\overline S}}\!\!q\left({r_{l,j}|\left[{{\bf x}_{k,j}}\right]_i;\sigma_w^2}\right)~p\left({\left[{{\bf x}_{k,j}}\right]_i;a_k}\right).
\end{align}

Then, the {\it a posteriori} mean and variance of $\left[{{\bf x}_{k,j}}\right]_i$, $\forall i,j,k$, are given, respectively, as
\begin{align}
\label{eq:postmeanAMP} {\left[{\widehat{\bf x}_{k,j}}\right]_i}&=f_m\left({r_{l,j},\phi_{l,j}}\right)\nonumber\\
&=\sum\limits_{\left[{{\bf x}_{k,j}}\right]_i\in\mathbb{\overline S}}\left[{{\bf x}_{k,j}}\right]_i~q\left({\left[{{\bf x}_{k,j}}\right]_i|r_{l,j}, \phi_{l,j};\sigma_w^2, a_k}\right),\\
\label{eq:postvarAMP}{\left[{{\widehat{\bf v}}_{k,j}}\right]_i}&=f_v\left({r_{l,j},\phi_{l,j}}\right)\nonumber\\
&=\!-\left|{\left[{\widehat{{\bf x}}_{k,j}}\right]_i}\right|^2\!\!\!+\!\!\!\!\!\!\!\sum\limits_{\left[{{\bf x}_{k,j}}\right]_i\in\mathbb{\overline S}}\!\!\!\!\!\!\left|{\left[{{\bf x}_{k,j}}\right]_i}\right|^2\!\!q\!\!\left({\left[{{\bf x}_{k,j}}\right]_i\!|r_{l,j}, \phi_{l,j};\sigma_w^2, a_k}\right),
\end{align}
where $l=(k-1)N_t+i$.

It is worth noting that $r_{l,j}$, $\phi_{l,j}$, $\widehat{{\bf x} }_{k,j}$, and $\widehat{{\bf v} }_{k,j}$ are updated iteratively by the AMP algorithm. In the factor graph in Fig. \ref{fig:FactorGraph}, in particular, $r_{l,j}$ and $\phi_{l,j}$, $\forall l,j$, are updated iteratively at the variable nodes. The update rules at the $t$-th iteration are expressed as
\begin{align}
\label{eq:UpdateSigma} \phi_{l,j}^t&=\left({\sum\limits_{n=1}^{N_r}\dfrac{\left|{\bf H}_{[n,l]}\right|^2}{\sigma_w^2+V_{n,j}^{t}}}\right)^{-1},\\
\label{eq:UpdateR} r_{l,j}^t&=\left[{\widehat{{\bf x}}_{k,j}^t}\right]_i+\phi_{l,j}^t\sum\limits_{n=1}^{N_r}\dfrac{{\bf H}^*_{[n,l]}\left({\left[{{\bf y}_j}\right]_n-Z_{n,j}^t}\right)}{\sigma_w^2+V_{n,j}^{t}},
\end{align}
where $V_{n,j}^{t}$ and $Z_{n,j}^t$, $\forall n,j$, are updated at the factor nodes of the factor graph as
\begin{align}
\label{eq:UpdateV} V_{n,j}^t&=\sum\limits_{k=1}^K\left|{{\bf H}_k}_{[{n,:}]}\right|^2\widehat{{\bf v} }_{k,j}^t,\\
\label{eq:UpdateZ} Z_{n,j}^t&=\sum\limits_{k=1}^K{{\bf H}_k}_{[{n,:}]}\widehat{{\bf x} }_{k,j}^t-V_{n,j}^t\dfrac{\left[{{\bf y}_j}\right]_n-Z_{n,j}^{t-1}}{\sigma_w^2+V_{n,j}^{t-1}}.
\end{align}

Due to space limitations, interested readers are referred to \cite{AMPmeng,KeMaLongTSP} for the detailed derivation of the AMP update rules in (\ref{eq:UpdateSigma})$-$(\ref{eq:UpdateZ}).

\subsubsection{Parameters Estimation}
It is noteworthy that the activity indicator $a_k$, $\forall k\in[K]$, and the noise variance $\sigma_w^2$ in (\ref{eq:postmeanAMP}) and (\ref{eq:postvarAMP}) are unknown parameters to be determined. Hence, we will integrate the EM algorithm to adaptively learn those parameters in each DS-AMP iteration.

Note that the EM algorithm is an iterative approach that finds the maximum likelihood solutions for probabilistic models with unknown parameters \cite{EM}. By defining ${\bm \theta}=\left\{{\sigma_w^2, a_k, k\in[K]}\right\}$, the EM algorithm updates the parameter set ${\bm \theta}$ as follows
\begin{align}
\label{eq:EM1} Q\left({{\bm \theta},{\bm \theta}^t}\right)&=\mathbb{E}\left\{{{\rm ln}~p\left({{\bf X, Y;{\bm \theta}}}\right)|{\bf Y};{\bm \theta}^t}\right\},\\
\label{eq:EM2} {\bm \theta}^{t+1}&={\rm arg}\max\limits_{{\bm \theta}}Q\left({{\bm \theta},{\bm \theta}^t}\right),
\end{align}
where ${\bm \theta}^t$ is the parameter set estimated at the $t$-th iteration, $\mathbb{E}\left\{{\cdot|{\bf Y};{\bm \theta}^t}\right\}$ denotes the expectation conditioned on the received signal ${\bf Y}$ under ${\bm \theta}^t$.

According to the AMP algorithm, the {\it a posteriori} distribution $p\left({\tilde{\bf x}_j|{\bf y}_j;\sigma_w^2, {\bf a}}\right)$ can be approximated as $q\left({\tilde{\bf x}_j|{\bf y}_j;\sigma_w^2, {\bf a}}\right)$, which reduces the complexity of the EM estimation. Hence, we can obtain the update rules of the noise variance $\sigma_w^2$ and the activity indicator $a_k$, $\forall k$, as follows
\begin{align}
\label{eq:EMnoiseVar}\left({\sigma_w^2}\right)^{t+1}&=\dfrac{1}{JN_r}\sum\limits_{j=1}^{J}\sum\limits_{n=1}^{N_r}\left[{\dfrac{\left({\left[{{\bf y}_j}\right]_n-Z_{n,j}^t}\right)^2}{\left({1+\frac{V_{n,j}^t}{\left({\sigma_w^2}\right)^t}}\right)^2}+\dfrac{\left({\sigma_w^2}\right)^t V_{n,j}^t}{V_{n,j}^t+\left({\sigma_w^2}\right)^t}}\right],\\
\label{eq:EMactivityIndica} a_k^{t+1}&=f_a\left({r_{l,j}^t,\phi_{l,j}^t;a_k^t}\right)\nonumber\\
&=\dfrac{1}{J}\sum\limits_{j=1}^{J}\sum\limits_{{\bf x}_{k,j}\in \Gamma_0}\prod\limits_{i=1}^{N_t}q\left({\left[{{\bf x}_{k,j}}\right]_i|r_{l,j}^t, \phi_{l,j}^t;a_k^t}\right),
\end{align}
where $l=(k-1)N_t+i$ and $\Gamma_0$ is the set of all possible ${\bf x}_{k,j}$ when the $k$-th MTD is active\footnote{For example, if $N_t=2$ and $\mathbb{S}=\left\{{+1,-1}\right\}$, then
\begin{equation}\label{eq:MediaModulated_SignalSet}
{\Gamma_0} = {\left\{ {\left[ {\begin{array}{*{10}{c}}
+1\\
0
\end{array}} \right],\left[ {\begin{array}{*{10}{c}}
-1\\
0
\end{array}} \right],\left[ {\begin{array}{*{10}{c}}
0\\
+1
\end{array}} \right],\left[ {\begin{array}{*{10}{c}}
0\\
-1
\end{array}} \right]} \right\}}.
\end{equation}}.
The detailed derivation of the EM update rules as indicated in (\ref{eq:EMnoiseVar}) and (\ref{eq:EMactivityIndica}) is available in the Appendix.

\subsubsection{Arithmetic Flow of the Proposed DS-AMP Algorithm}
Based on (\ref{eq:PosMarginalAMP})$-$(\ref{eq:EMactivityIndica}), we summarize our proposed DS-AMP algorithm in {\bf Algorithm \ref{Algorithm:1}}.
\SetAlgoNoLine
\SetAlFnt{\small}
\SetAlCapFnt{\normalsize}
\SetAlCapNameFnt{\normalsize}\
\begin{algorithm}[!t]
\caption{Proposed DS-AMP Algorithm}\label{Algorithm:1}
\begin{algorithmic}[1]
\REQUIRE The received signals ${\bf{Y}}\!\!=\!\!\left[{{\bf {y}}_{1}, ..., {\bf {y}}_{J}}\right]\!\in\!\mathbb{C}^{N_r \!\times\! J}$, the channel matrix ${{\bf{H}}}\!\!=\!\!\left[{{\bf H} _1,...,{\bf H} _K}\right] \!\in\!\mathbb{C}^{N_r \!\times \!(K N_t)}$, and the maximum iteration number $T_0$.
\ENSURE The set of active MTDs $\Omega$ and the reconstructed media modulation signal ${\bf X}\in\mathbb{C}^{KN_t \!\times\! J} $.
\STATE ${\forall i,j,k,n}$: We initialize the iterative index $t$=1, the activity indicator $a^1_k\!=\!0.5$, $Z^0_{n,j}\!=\!\left[{{\bf y}_j}\right]_n$, $V^0_{n,j}\!=\!1$, the noise variance $\left({\sigma_w^2}\right)^1=100$, the reconstructed signal ${\bf X}={\bf 0}_{KN_t\times J}$, $\left[{\widehat{{\bf x}}_{k,j}^1}\right]_i\!\!=\!\!a^1_k\sum\limits_{s\in\mathbb S} s/MN_t$, and $\left[{\widehat{{\bf v}}_{k,j}^1}\right]_i=a^1_k\sum\limits_{s\in\mathbb S} \left|{s}\right|^2/MN_t-\left|{\left[{\widehat{{\bf x}}_{k,j}^1}\right]_i}\right|^2$;
\label{A1:initial}
\FOR {$t=1$ to $ T_0$}
\label{A1:T0}
\STATE \textbf{\textbf{\%}AMP operation:}
\STATE ${\forall i,j,k,n}$: Compute $V_{n,j}^t$, $Z_{n,j}^t$, $\phi_{l,j}^t$, and $r_{l,j}^t$ by using (\ref{eq:UpdateV}), (\ref{eq:UpdateZ}), (\ref{eq:UpdateSigma}), and (\ref{eq:UpdateR}), respectively, where $l\!\!=\!\!(k-1)N_t\!+\!i$;~~\{Decoupling step\}
\label{A1:decoupling}
\STATE ${\forall i,j,k,n}$: Compute $\left[{\widehat{{\bf x}}_{k,j}^{t+1}}\right]_i$ and $\left[{\widehat{{\bf v}}_{k,j}^{t+1}}\right]_i$ by using (\ref{eq:postmeanAMP}) and (\ref{eq:postvarAMP}), respectively;~~\{Denoising step\}
\label{A1:denoising}
\STATE \textbf{\textbf{\%}EM operation:}
\STATE ${\forall k}$: Compute $(\sigma_w^2)^{t+1}$ and $a^{t+1}_k$ by using (\ref{eq:EMnoiseVar}) and (\ref{eq:EMactivityIndica});
\label{A1:EM}
\ENDFOR
\label{A1:endfor}
\STATE \textbf{\textbf{\%}Min-max normalization:}
\STATE Let $\tilde{{\bf a}}\!=\!\!\frac{\widehat{\bf a}-{\rm min}(\widehat{\bf a})}{{\rm max}(\widehat{\bf a})-{\rm min}(\widehat{\bf a})}$, where $\widehat{\bf a}\!\!=\!\!\left[{\widehat{a}_1,...,\widehat{a}_K}\right]^T\!\!=\!\!\left[{a_1^{T_0},...,a_K^{T_0}}\right]^T$, ${\rm min}(\cdot)$ and ${\rm max}(\cdot)$ are the minimum value and maximum value of the arguments, respectively;
\label{A1:minmax}
\STATE \textbf{\textbf{\%}Extract the active MTDs and their MAPs:}
\STATE ${\forall k}$: The set of active MTDs $\Omega=\{k|\left[{\tilde{{\bf a}}}\right]_k>0.5\}$;
\label{A1:AUD}
\STATE ${\forall k,j}$: $\eta^*\!=\!{\rm arg\mathop{max}\nolimits}_{\widehat{\eta}\in [N_t]}\left[{\widehat{\bf x}_{k,j}^{T_0}}\right]_{\widehat\eta}$;
\label{A1:MAPs}
\STATE ${\forall k\in\Omega,\forall j}$: \\
The reconstructed signal is ${\bf X}_{\left[{(k-1)N_t+\eta^*,j}\right]}=\left[{\widehat{\bf x}_{k,j}^{T_0}}\right]_{\eta^*}$.
\label{A1:signalreconst}
\end{algorithmic}
\end{algorithm}

The details of {\bf Algorithm \ref{Algorithm:1}} are explained as follows. Since the active/inactive status of the MTDs is unknown, we initialize the activity indicator $a^1_k=0.5$, $\forall k\in[K]$, in line \ref{A1:initial}. We note that lines \ref{A1:decoupling}$-$\ref{A1:denoising} are the key steps of the AMP operation, which consists of two parts. For the first part, a decoupling operation is performed in the same way as the original AMP algorithm in order to decouple the superimposed received signal into uncoupled scalar elements based on (\ref{eq:systemdecoupled}) and (\ref{eq:UpdateSigma})$-$(\ref{eq:UpdateZ}) \cite{AMP,AMPmeng,KeMaLongTSP}. For the second part, the denoising step computes the {\it a posteriori} mean and variance of each scalar element by using (\ref{eq:postmeanAMP}) and (\ref{eq:postvarAMP}), where the structured sparsity in the modulation domain and the discrete distribution of the QAM alphabet are exploited in the {\it a priori} probability. In line \ref{A1:EM}, based on (\ref{eq:EMnoiseVar}) and (\ref{eq:EMactivityIndica}), the EM algorithm updates the noise variance $\sigma_w^2$ and the activity indicators $a_k$, $\forall k\in[K]$, by exploiting the structured sparsity in the time domain. Furthermore, the iteration stops when a predefined maximum number $T_0$ is reached, where $T_0$ is chosen to guarantee the convergence of the algorithm. Line \ref{A1:minmax} linearly transforms the estimated activity indicator vector $\widehat{\bf a}$ to $\tilde{\bf a}$ by using the min-max normalization. In line \ref{A1:AUD}, if the $k$-th ($k\in[K]$) element of $\tilde{\bf a}$ is larger than 0.5, the $k$-th MTD is considered to be active. Line \ref{A1:MAPs} selects the possible MAPs based on the structured sparsity of the media modulated symbol as indicated in (\ref{eq:d}). Finally, line \ref{A1:signalreconst} reconstructs the media modulated signals.

Note that all the estimated activity indicators can be small, i.e., $[\widehat{\bf a}]_k<0.5, \forall k\in[K]$, in poor conditions, e.g., at low signal-to-noise ratio (SNR), where the set of active MTDs given in \cite{AMP-AUD} (i.e., $\{k|[\widehat{\bf a}]_k>0.5\}$, $\forall k\in[K]$) results in the degraded active MTDs detection error. Fortunately, we find that the elements of $\widehat{\bf a}$ are linearly separable in most cases, even when all the elements are smaller than 0.5. Hence, in line \ref{A1:minmax}, we preprocess $\widehat{\bf a}$ by using the min-max normalization to enlarge the separability of its elements, and the output is $\tilde{\bf a}$. After the preprocessing, in line \ref{A1:AUD}, we can obtain the set of active MTDs based on $\tilde{\bf a}$. The simulation results illustrated in Section V verify the advantage of our improved active MTDs detection method over that in \cite{AMP-AUD}.

\vspace{-3.8mm}
\subsection{State Evolution of the DS-AMP Algorithm}
SE is a tool for analyzing the performance of AMP algorithms in the large system limit, i.e., $KN_t\rightarrow \infty$, by tracking the mean-square errors (MSE) of each iteration \cite{AMP,KeMaLongTSP}. In particular, capitalizing on the SE, we can characterize the performance of the proposed DS-AMP algorithm theoretically.

To start with, the MSE and average variance of the estimated signals at the $t$-th iteration are respectively defined as
\begin{align}
\label{eq:SEMSE} e^t&=\dfrac{1}{KJN_t}\sum\limits_{k=1}^{K}\sum\limits_{j=1}^{J}\sum\limits_{i=1}^{N_t}\left|{[\widehat{\bf x}_{k,j}^t]_i-[{\bf x}_{k,j}]_i}\right|^2,\\
\label{eq:SEVariance} v^t&=\dfrac{1}{KJN_t}\sum\limits_{k=1}^{K}\sum\limits_{j=1}^{J}\sum\limits_{i=1}^{N_t}[\widehat{\bf v}_{k,j}^t]_i.
\end{align}

We define a scalar random variable $X_0$ that obeys the {\it a priori} distribution in (\ref{eq:PriorMarginal}). Based on \cite{KeMaLongTSP}, it can be shown, in large system limit and if the elements of the measurement matrix follow an i.i.d. distribution with zero mean and variance $\gamma$, that the estimated mean of $x_0$ at the $t$-th iteration, denoted as $r_0^t$, can be expressed as
\begin{equation}\label{eq:SEupdateR}
\begin{array}{l}
r_0^t=x_0+\sqrt{\dfrac{\sigma_w^2+\gamma K N_t e^t}{N_r\gamma}}z,
\end{array}
\end{equation}

where $x_0$ is a realization of $X_0$, $z\sim{\cal C}{\cal N}(z;0,1)$. In addition, the estimated variance of $x_0$ at the $t$-th iteration, denoted as $\phi_0^t$, can be expressed as
\begin{equation}\label{eq:SEupdateSigma}
\begin{array}{l}
\phi_0^t\approx \dfrac{\sigma_w^2+\gamma KN_t v^t}{N_r\gamma}.
\end{array}
\end{equation}

By substituting (\ref{eq:SEupdateR}) and (\ref{eq:SEupdateSigma}) into (\ref{eq:postmeanAMP}) and (\ref{eq:postvarAMP}), we obtain the approximated MMSE estimation of $x_0$ for the DS-AMP algorithm at the $(t+1)$-th iteration. Hence, the MSE and the average variance of the estimated signals at the $(t+1)$-th iteration can be expressed as
\begin{align}
\label{eq:SEMSEUpdate} e^{t+1}&=\int dx_0p_0(x_0)\int{\cal D}z\left|{f_m(r_0^t,\phi_0^t)-x_0}\right|^2,\\
\label{eq:SEVarianceUpdate} v^{t+1}&=\int dx_0p_0(x_0)\int{\cal D}zf_v(r_0^t,\phi_0^t),
\end{align}
where $p_0(x_0)$ is the {\it a priori} distribution in (\ref{eq:PriorMarginal}), ${\cal D}z=e^{-|z|^2}/\pi dz$, $f_m(r_0^t,\phi_0^t)$ and $f_v(r_0^t,\phi_0^t)$ are defined in (\ref{eq:postmeanAMP}) and (\ref{eq:postvarAMP}), respectively.

As listed in {\bf Algorithm} {\ref{Algorithm:2}}, we present the detailed procedures of the SE of the proposed DS-AMP algorithm. In particular, we adopt Monte Carlo simulations to generate a large number of realizations of the transmit signals, where the sporadic traffic and the doubly structured sparsity are fully embodied. Simulation results in Section V-B will demonstrate that the SE can accurately predict the simulated results of the proposed DS-AMP algorithm.

\SetAlgoNoLine
\SetAlFnt{\small}
\SetAlCapFnt{\normalsize}
\SetAlCapNameFnt{\normalsize}\
\begin{algorithm}[!t]
\caption{State Evolution of DS-AMP Algorithm}\label{Algorithm:2}
\begin{algorithmic}[1]
\REQUIRE The noise variance $\sigma_w^2$, the sparsity level $\lambda=\frac{K_a}{K}$, the number of MAPs $N_t$, the frame length $J$, the order of the QAM modulation, the variance $\gamma$ of the elements in the measurement matrix, the number of Monte Carlo simulations $N_{\rm MC}$, the maximum SE iterations $T_{\rm SE}$, and the terminal threshold $\varepsilon$.
\ENSURE The theoretically predicted MSE $\widehat{e}$.
\STATE $\forall m\in[N_{\rm MC}]$: Generate $N_{\rm MC}$ realizations of the transmit signals ${\bf X}^m\in\mathbb{C}^{KN_t \times J}$,  according to the {\it a priori} distribution in (\ref{eq:PriorGeneral}).
\STATE $\forall m,k$: Define ${\bf e}^1={\bf 0}_{N_{\rm MC}\times 1}$ and ${\bf v}^1={\bf 0}_{N_{\rm MC}\times 1}$ to record the predicted MSE and average variance of the $m$-th Monte Carlo realization. We initialize the iteration number $t=1$, the predicted MSE $e^1=1$, the average variance $v^1=1$, and the activity indicators for the $m$-th signal realization $a_{k,m}^1=0.5$;
\label{A2:initial}
\FOR{$t=1$ to $T_{\rm SE}$}
\FOR {$m=1$ to $N_{\rm MC}$}
\label{A1:TSE}
\STATE $\forall i,j,k$: \!$r_{l,j}^{m,t}\!\!=\!\!\!\left[{{\bf x}_{k,j}^{m}}\right]_i\!\!\!+\!\!\sqrt{\frac{\sigma_w^2+\gamma K N_t e^t}{N_r\gamma}}z$, \!$\phi_{l,j}^{m,t}\!\!=\!\!\frac{\sigma_w^2+\gamma KN_t v^t}{N_r\gamma}$;
\STATE $\forall i,j,k$: \!\!$\left[{\widehat{\bf x}_{k,j}^{m}}\right]_i\!\!=\!\!f_m(r_{l,j}^{m,t},\phi_{l,j}^{m,t})$, $\left[{\widehat{\bf v}_{k,j}^{m}}\right]_i\!\!=\!\!f_v(r_{l,j}^{m,t},\phi_{l,j}^{m,t})$;
\label{Algo2:PostX}
\STATE $\forall k$: $a_{k,m}^{t+1}=f_a(r_{l,j}^{m,t},\phi_{l,j}^{m,t};a_{k,m}^{t})$;
\STATE Calculating $\left[{{\bf e}^{t+1}}\right]_m$ and $\left[{{\bf v}^{t+1}}\right]_m$ referring to (\ref{eq:SEMSE}) and (\ref{eq:SEVariance}), respectively;
\ENDFOR
\STATE $e^{t+1}\!\!=\!\!\frac{1}{N_{\rm MC}}\sum\nolimits_{m=1}^{N_{\rm MC}}\left[{{\bf e}^{t+1}}\right]_m$, $v^{t+1}\!\!=\!\!\frac{1}{N_{\rm MC}}\sum\nolimits_{m=1}^{N_{\rm MC}}\left[{{\bf v}^{t+1}}\right]_m$;
\STATE $\widehat{e}=e^{t+1}$;
\IF {$\left|{e^{t+1}-e^t}\right|<\varepsilon$}
\STATE break;~~\{End the SE iterations\}
\ENDIF
\ENDFOR
\end{algorithmic}
\end{algorithm}

\vspace{-6mm}
\subsection{Computational Complexity of the DS-AMP Algorithm}
The computational complexity of the proposed DS-AMP algorithm mainly arises from the complex-valued matrix multiplications\footnote{For the sake of simplicity, a real-valued multiplication is assumed to have a complexity that equals a quarter of a complex-valued multiplication.} of the following operations at each iteration \cite{AMP-SM}.


{\bf AMP decoupling step}: The complexity of performing the AMP decoupling step, i.e., (\ref{eq:UpdateSigma})$-$(\ref{eq:UpdateZ}), is $\mathcal{O}(\frac{5}{2}JKN_tN_r)$.

{\bf AMP denoising step}: The complexity of performing the AMP denoising step, i.e., (\ref{eq:NormulizeItem})$-$(\ref{eq:postvarAMP}), is $\mathcal{O}\left[{JKN_t(|\mathbb{S}|_c+\frac{1}{4})}\right]$.

Furthermore, simulation results demonstrate that the predefined maximum number of iterations $T_0$ can be small to guarantee the convergence of the proposed DS-AMP algorithm. Hence the overall complexity is of the order of ${\cal O}\left[{T_0JKN_t(\frac{5}{2}N_r+|\mathbb{S}|_c+\frac{1}{4})}\right]$, which scales linearly with the number of MTDs, the number of MAPs in media modulation, the order of the QAM modulation, and the number of receive antennas at the BSs. This linear complexity is appealing for efficiently processing the massive access of future IoT.

\begin{figure*}[t]
\vspace{-8mm}
     \centering
     \includegraphics[width=14cm, keepaspectratio]%
     {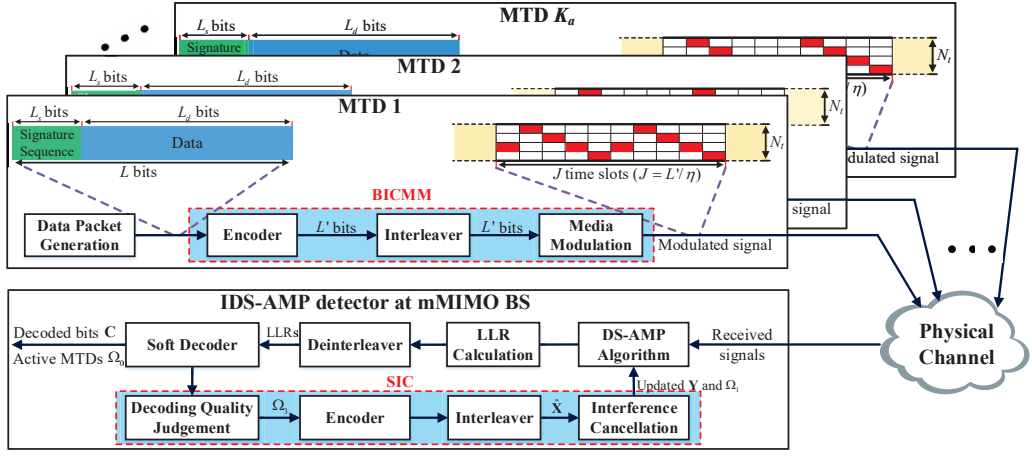}
     \captionsetup{font={footnotesize}, singlelinecheck = off, justification = raggedright, name={Fig.},labelsep=period}
     \caption{Communication process of the proposed massive access solution for coded media modulation based mMTC.}
     \label{fig:Communication}
     \vspace*{-5mm}
\end{figure*}

\vspace{-3.9mm}
\section{Proposed Solution for Coded Media Modulation Based mMTC}
In this section, we propose a solution for coded media modulation based mMTC scheme, whose block diagram is illustrated in Fig. \ref{fig:Communication}. Specifically, we propose a dedicated data packet structure as well as a BICMM scheme at the MTDs, and develop an IDS-AMP detector at the receiver, so that the DADD performance can be further improved.

\vspace{2mm}
\subsection{Dedicated Data Packet Structure and BICMM at the MTDs}
\vspace{-1.5mm}
The proposed data packet structure is tailored for the implementation of an SIC-based IDS-AMP detector at the receiver. Specifically, the data packet is composed of two parts, i.e., the $L_s$-bit signature sequence part and the $L_d$-bit payload data. We consider that all the MTDs share the same binary signature sequence ${\bf b}_s\in \mathbb{N}^{L_s\times 1}$, which is a pre-defined pseudo-random 0/1 sequence known at the transceiver. At the receiver, we consider the Hamming distance $D({\bf b}_s,\widehat{\bf b}_s)$ between ${\bf b}_s$ and $\widehat{\bf b}_s$ as a metric to evaluate the decoding quality of the associated payload data part of the MTDs, where $\widehat{\bf b}_s\in \mathbb{N}^{L_s\times 1}$ is the estimated binary signature sequence of any detected MTD. Hence, $D({\bf b}_s,\widehat{\bf b}_s)$ can be used to determine the order for the subsequent SIC processing. By contrast, the conventional SIC order metric, e.g., the SNR or the signal-to-interference-plus-noise ratio (SINR), is difficult to be acquired in practical massive access processing \cite{MIMO-OFDM}. Particularly, for $0\leq D({\bf b}_s,\widehat{\bf b}_s)\leq L_s$, a smaller $D({\bf b}_s,\widehat{\bf b}_s)$ indicates a better decoding quality of the signature sequence, which usually implies a higher decoding quality of the associated payload data and a higher priority to be subtracted during the SIC stage. As a result, the error propagation in the SIC processing can be mitigated with the aid of the proposed signature sequence.

In addition to the data packet structure, we develop a BICMM scheme. As shown in Fig. \ref{fig:Communication}, the BICMM scheme consists of an encoder, a bit-wise interleaver, and a media modulation module. Specifically, after channel coding, the length of one data packet is expanded from $L$ bits to $L'$ bits. Then, the $L'$-bit data packet is delivered to a bit-wise interleaver module. We consider a block interleaver with $\eta$ columns and $J=L'/\eta$ rows, where we assume $L'=\eta J$ without loss of generality. The $L'$-bit data packet is fed into the interleaver by rows and read out by columns\footnote{Note that the deinterleaver is an array identical to the interleaver, while the $L'$ bits data packet is read in by columns and read out by rows.}. At the transmitter, every $\eta$ bits of the interleaved $L'$-bit data packet are sequentially modulated into $J$ media modulation symbols that are transmitted in $J$ successive time slots (i.e., a frame). Hence, after interleaving, the bits originally associated with the same media modulated symbol are separated by ($J-1$) other bits, which makes the coded bits to the media modulation module more dispersed to combat the spatial-selective channel fading\footnote{Practical slowly time-varying fading channels may exhibit burst error which is beyond the error correction capability of the codes of reasonable complexity\cite{GoldSmith}. Therefore, in fading channels, coding is typically combined with interleaving to mitigate the effect of burst error\cite{GoldSmith}.}. 
In particular, each media modulation-based MTD has $N_t$ unique radiation patterns associated with $N_t$ different channels with different spatial channel fading. Typically, active MTD selects one of the channels based on its input spatial data bits to transmit a media modulated symbol with $\eta$ bits. Burst errors may occur if the selected channel suffers from deep fading, which results in the spatial-selective channel fading.
Thanks to the bit-wise interleaver combined with the encoder, $\eta$ successive bits originally associated with the same media modulated symbol are separated into different media modulated symbols with different spatial channel fading, which mitigates the spatial-selective channel fading and improves the data decoding performance \cite{BICM,BICSM,GoldSmith}.

\vspace{-6mm}
\subsection{Proposed IDS-AMP Detector at the BS}
As shown in Fig. \ref{fig:Communication}, the proposed IDS-AMP detector has 8 modules, including a DS-AMP algorithm module, an LLR calculation module, a deinterleaver, a soft decoder, a decoding quality judgement module, an encoder, an interleaver, and an interference cancellation module. The main procedure is summarized in {\bf Algorithm \ref{Algorithm:3}}, and the details are illustrated as follows.

{\bf DS-AMP algorithm module} (lines \ref{A3:JSAMP}$-$\ref{A3:OMEGA2} of {\bf Algorithm \ref{Algorithm:3}}): Firstly (line {\ref{A3:JSAMP}}), we obtain the approximated {\it a posteriori} distribution $q\left({{\bf x}_{k,j}|{\bf y}_j;\sigma_w^2,a_k}\right)$, $\forall k,j$, and the activity indicator vector $\widehat{\bf a}=[\widehat{a}_1,...,\widehat{a}_K]^T$, which can be acquired by calling the DS-AMP algorithm (i.e., performing lines \ref{A1:initial}$-$\ref{A1:endfor} of {\bf Algorithm \ref{Algorithm:1}}). Secondly (lines \ref{A3:IfI=1}$-$\ref{A3:endIfI=1}), if the iteration index is $i=0$, we acquire the indices of the active MTDs that are detected, denoted as $\Omega_0$, by executing lines \ref{A1:minmax}$-$\ref{A1:AUD} of {\bf Algorithm \ref{Algorithm:1}}. The index set of the MTDs that remain to be iteratively decoded, denoted by $\Omega_1$, is assigned to be equivalent to $\Omega_0$ in the first iteration (i.e., the iteration with index $i=0$) and is updated in the subsequent SIC. Thirdly (lines \ref{A3:updateI}, \ref{A3:OMEGA2}), we update the iteration index $i=i+1$, and then select the $\overline{N}$ MTDs most likely to be active based on the activity indicators for the subsequent SIC, where the index set of the $\overline{N}$ MTDs is denoted as $\Omega_2\!\!=\!\!\Theta([\widehat{\bf a}]_{\Omega_1},\!\overline{N})$ and we consider the MTD with a larger activity indicator is more likely to be active. If $\left|{\Omega_1}\right|_c$ is smaller than the predefined constant $\overline{N}$, we set $\Omega_2=\Omega_1$.

{\bf LLR calculation module} (line \ref{A3:LLR1} of {\bf Algorithm \ref{Algorithm:3}}): According to the approximated {\it a posteriori} distribution $q\left({{\bf x}_{k,j}|{\bf y}_j;\sigma_w^2,a_k}\right)$, $\forall k,j$, obtained from the DS-AMP algorithm module, we calculate the LLR information of the MTDs whose indices are in $\Omega_2$. For any media modulation symbol ${\bf x}_{k,j}$, $\forall k,j$, the LLR of the associated media modulated bit $B_{k,j,b}^{\rm MED}$, $\forall b\in[{\rm log}_2N_t]$, and the LLR of the associated $M$-QAM bit $B_{k,j,d}^{\rm QAM}$, $\forall d\in[{\rm log}_2M]$, can be expressed, respectively, as
\begin{align}
\label{eq:LLR1}{\rm LLR}\left({B_{k,j,b}^{\rm MED}}\right)&={\rm log}\dfrac {\sum\nolimits_{{\bf x}_{k,j}\in\Phi_0^{b}} q\left({{\bf x}_{k,j}|{\bf y}_j;\sigma_w^2,a_k}\right)}{\sum\nolimits_{{\bf x}_{k,j}\in\Phi_1^{b}}q\left({{\bf x}_{k,j}|{\bf y}_j;\sigma_w^2,a_k}\right)},\\
\label{eq:LLR2}{\rm LLR}\left({B_{k,j,d}^{\rm QAM}}\right)&={\rm log}\dfrac {\sum\nolimits_{{\bf x}_{k,j}\in\Psi_0^{d}} q\left({{\bf x}_{k,j}|{\bf y}_j;\sigma_w^2,a_k}\right)}{\sum\nolimits_{{\bf x}_{k,j}\in\Psi_1^{d}}q\left({{\bf x}_{k,j}|{\bf y}_j;\sigma_w^2,a_k}\right)},
\end{align}
where $\Phi_0^b$ ($\Phi_1^b$) is the set of ${\bf x}_{k,j}$ for which the media modulated bit $B_{k,j,b}^{\rm MED}$, $\forall b$, is equal to zero (one), and $\Psi_0^d$ ($\Psi_1^d$) is the set of ${\bf x}_{k,j}$ for which the $M$-QAM bit $B_{k,j,d}^{\rm QAM}$, $\forall d$, is equal to zero (one)\footnote{For example, supposing that $N_t=2$, $M=2$, $\mathbb{S}=\{+1,-1\}$, $b\in[1]$, and $d\in[1]$, then we can get
\begin{equation}\label{eq:examplePhi}
\Phi_0^1 = {\left\{ {\left[ {\begin{array}{*{10}{c}}
+1\\
0
\end{array}} \right],\left[ {\begin{array}{*{10}{c}}
-1\\
0
\end{array}} \right]} \right\}},
\Phi_1^1 = {\left\{ {\left[ {\begin{array}{*{10}{c}}
0\\
+1
\end{array}} \right],\left[ {\begin{array}{*{10}{c}}
0\\
-1
\end{array}} \right]} \right\}},
\end{equation}

\begin{equation}\label{eq:examplePsi}
\Psi_0^1 = {\left\{ {\left[ {\begin{array}{*{10}{c}}
+1\\
0
\end{array}} \right],\left[ {\begin{array}{*{10}{c}}
0\\
+1
\end{array}} \right]} \right\}},
\Psi_1^1 = {\left\{ {\left[ {\begin{array}{*{10}{c}}
-1\\
0
\end{array}} \right],\left[ {\begin{array}{*{10}{c}}
0\\
-1
\end{array}} \right]} \right\}}.
\end{equation}}.

{\bf Deinterleaver and soft decoder modules} (line {\ref{A3:DEINTER1-DECODE1}} of {\bf Algorithm \ref{Algorithm:3}}): Firstly, the LLR information of the MTDs with indices in $\Omega_2$ is deinterleaved. Note that the deinterleaver is an array identical to the interleaver described in Section IV-A, with the exception that the $L'$-bit data packet is read in by columns and read out by rows. Secondly, for each MTD whose index is in $\Omega_2$, the soft decoder decodes the deinterleaved LLR information to get the data packet of $L_s+L_d$ bits. Finally, for the $|\Omega_2|_c$ MTDs detected as active, the decoded $L|\Omega_2|_c$ bits are recorded in a binary data packet matrix ${\bf B}\in\mathbb{N}^{K\times L}$, i.e., ${\bf B}_{[\Omega_2,1:L]}$. In particular, the first $L_s$ columns of ${\bf B}$ store the decoded signature sequence bits and the remaining $L_d$ columns record the decoded payload data bits.

{\bf Decoding quality judgement module} (lines \ref{A3:DQJ}$-${\ref{A3:WeightDecision} of {\bf Algorithm \ref{Algorithm:3}}): Firstly (lines \ref{A3:weight}), for each MTD whose index is in $\Omega_2|_{\overline{n}}$, $\forall \overline{n}\in[|\Omega_2|_c]$, we calculate the Hamming distances between the decoded signature sequence $\left({\bf B}_{[\Omega_2|_{\overline{n}},1:L_s]}\right)^T$ and the true signature sequence ${\bf b}_s$, denoted as $[{\bf m}]_{\overline n}=D\left({{\bf b}_s,\left({\bf B}_{[\Omega_2|_{\overline{n}},1:L_s]}\right)^T}\right)$. Secondly (line \ref{A3:WeightDecision}), based on the Hamming distances ${\bf m}\in\mathbb{N}^{|\Omega_2|_c\times1}$, whether and how to perform the SIC processing is judged. In particular, if no element of ${\bf m}$ is equal to zero, we skip the interference cancellation module, decode the remaining MTDs indexed in $\{\Omega_1\setminus\Omega_2\}$ in line \ref{A3:LLR2}, and stop the algorithm in line {\ref{A3:break1}}. If there exist elements equal to zero in ${\bf m}$, on the other hand, this indicates that there are one or several MTDs with almost perfect decoding quality. In this case, we apply the signals reconstruction module and the interference cancellation module (i.e., lines \ref{A3:SICstart}$-$\ref{A3:endif}).

\SetAlgoNoLine
\SetAlFnt{\footnotesize}
\SetAlCapFnt{\normalsize}
\SetAlCapNameFnt{\normalsize}\
\begin{algorithm}[!t]
\caption{Proposed IDS-AMP Detector}\label{Algorithm:3}
\begin{algorithmic}[1]
\REQUIRE The received signals ${\bf{Y}}\!\!=\!\![{\bf {y}}_{1}, ..., {\bf {y}}_{J}]\!\in\!\mathbb{C}^{N_r \!\times\! J}$ and the channel matrix ${\bf H}={\bf H}_0\!\!=\!\![{\bf H} _1,...,{\bf H} _K] \!\in\!\mathbb{C}^{N_r \!\times \!(K N_t)}$.
\ENSURE The estimated index set of active MTDs $\Omega_0$, and the associated demodulated payload data bits ${\bf B}_{[\Omega_0,L_s+1:L]}$ with ${\bf B}\in\mathbb{N}^{K\times L}$ defined as the binary data packet matrix of all MTDs.
\STATE {\bf Initialization}: The iteration index $i$=0, the binary data packet matrix ${\bf B}={\bf 0}_{K\times L}$, and for possible active MTDs given their temporary index set $\Lambda$, their MAPs' index set is defined as $\widetilde\Lambda\!\!=\!\!\{\widetilde \Lambda_n\}_{n=1}^{|\Lambda|_c}$, where $\widetilde\Lambda_n\!\!=\!\!\{N_t(\Lambda|_n\!\!-\!\!1)\!+\!u\}_{u=1}^{N_t}$ is the MAPs' index set of the $n$-th MTD in $\Lambda$ for $ n\in[|\Lambda|_c]$\label{A3:initial};
\WHILE {$1$}
\STATE \textbf{\textbf{\%}DS-AMP algorithm module:}
\STATE Obtain $\widehat{\bf a}=[\widehat{a}_1,...,\widehat{a}_K]^T$ and $q\left({{\bf x}_{k,j}|{\bf y}_j;\sigma_w^2,a_k}\right)$, $\forall k,j$, by performing lines \ref{A1:initial}$-$\ref{A1:endfor} of {\bf Algorithm \ref{Algorithm:1}};\label{A3:JSAMP}
\IF{$i=0$}\label{A3:IfI=1}
\STATE Acquire the index set of active MTDs detected, denoted as $\Omega_0$, by performing lines \ref{A1:minmax}$-$\ref{A1:AUD} of {\bf Algorithm \ref{Algorithm:1}};\label{A3:AUD}
\STATE The index set of MTDs remaining to be iteratively decoded is denoted as $\Omega_1$, and $\Omega_1=\Omega_0$ in the first iteration;\label{A3:OMEGA1}
\ENDIF\label{A3:endIfI=1}
\STATE $i=i+1$;\label{A3:updateI}
\STATE  Acquire the $\overline{N}$ MTDs most likely to be active based on the quantities of activity indicators for the
following SIC, whose index set is denoted as $\Omega_2\!\!=\!\!\Theta([\widehat{\bf a}]_{\Omega_1},\!\overline{N})$;\label{A3:OMEGA2}
\STATE \textbf{\textbf{\%}LLR calculation module:}
\STATE Calculate the LLR information of the MTDs indexed by $\Omega_2$\label{A3:LLR1};
\STATE \textbf{\textbf{\%}Deinterleaver and soft decoder modules:}
\STATE For each MTD with index in $\Omega_2$, perform deinterleaving and soft decoding, and then record the decoded bits in the binary matrix ${\bf B}_{[\Omega_2,1:L]}$\label{A3:DEINTER1-DECODE1};
\STATE \textbf{\textbf{\%}Decoding quality judgement module:}\label{A3:DQJ}
\STATE For each decoded MTD with index in $\Omega_2$, calculate the Hamming distances between the estimated signature sequence and the true signature sequence, denoted as $[{\bf m}]_{\overline n}=D\left({{\bf b}_s,\left({\bf B}_{[\Omega_2|_{\overline n},1:L_s]}\right)^T}\right), {\overline n}\in[|\Omega_2|_c]$;\label{A3:weight}
\IF{$[{\bf m}]_{\overline n}\neq 0, \forall {\overline n}$\label{A3:WeightDecision}}
\STATE For each MTD indexed by $\{\Omega_1\setminus\Omega_2\}$, calculate the LLR information, perform deinterleaving and soft decoding, and then record the decoded bits in matrix ${\bf B}_{[\{\Omega_1\setminus\Omega_2\},:]}$\label{A3:LLR2};
\STATE {\bf break};\label{A3:break1}~~\{End the algorithm\};
\ELSE
\STATE \textbf{\textbf{\%}Signals reconstruction module:}
\STATE Acquire the MTDs with very high decoding quality for SIC, and the associated index set is denoted as $\Omega_3=\{\Omega_2|_{\overline n}\mid [{\bf m}]_{\overline n}=0, \forall {\overline n}\}$;\label{A3:SICstart}
\STATE For each MTD with index in $\Omega_3$, encode and interleave the decoded data bits ${\bf B}_{[\Omega_3,:]}$, then perform media modulation on the interleaved bits\label{A3:ENCODE};
\STATE $\hat{\bf X}={\bf 0}_{KN_t\times J}$, record the reconstructed signals in $\hat{\bf X}_{[\widetilde{\Omega_3},:]}$\label{A3:record3};
\STATE \textbf{\textbf{\%}Interference cancellation module:}
\STATE ${\bf Y}={\bf Y}-{\bf H}_0\hat{\bf X}$;~~\{Update the received signals\}\label{A3:update-Y}
\STATE $\Omega_1=\{\Omega_1\setminus \Omega_3\}$, $\Lambda=\{\Omega_0\setminus \Omega_1\}$;\label{A3:update-omega1}\label{A3:SICend}
\STATE ${\bf H}={{\bf H}_0}_{[:,\{[KN_t]\setminus\widetilde{\Lambda}\}]}$;~~\{Update the measurement matrix\}\label{A3:update-H}
\IF{$\Omega_1=\varnothing$\label{A3:NoneDecision}}
\STATE {\bf break};~~\{End the algorithm\}\label{A3:break2}
\ENDIF
\label{A3:endif}
\ENDIF
\ENDWHILE
\STATE {\bf Results}: The estimated index set of active MTDs is $\Omega_0$ and the associated demodulated payload data is ${\bf B}_{[\Omega_0,L_s+1:L]}$\label{A3:END}.
\end{algorithmic}
\end{algorithm}

{\bf Signals reconstruction module} (lines \ref{A3:SICstart}$-$\ref{A3:record3} of {\bf Algorithm \ref{Algorithm:3}}): Firstly (line \ref{A3:SICstart}), we acquire the indices of the zero elements in ${\bf m}$, denoted by $\Omega_3=\{\Omega_2|_{\overline n}\mid [{\bf m}]_{\overline n}=0, \forall {\overline n}\}$, which is a fine-grained index set of the MTDs with higher decoding quality. Secondly (lines {\ref{A3:ENCODE}}), for each MTD with indices in $\Omega_3$, to reconstruct the signal components, we sequentially perform encoding, interleaving, and media modulation (i.e., we repeat the BICMM scheme at MTD) according to the decoded ${\bf B}_{[\Omega_3,1:L]}$. Thirdly (line \ref{A3:record3}), we record the reconstructed signals in $\hat{\bf X}_{[\widetilde{\Omega_3},:]}$, where $\widetilde{\Omega_3}$ is the index set of the MAPs of the MTDs whose indices are in $\Omega_3$.

{\bf Interference cancellation module} (lines \ref{A3:update-Y}$-$\ref{A3:endif} of {\bf Algorithm \ref{Algorithm:3}}): Firstly (line {\ref{A3:update-Y}}), we subtract the signal components $\hat{\bf X}_{[\widetilde{\Omega_3},:]}$ from the received signals ${\bf Y}$. Secondly (line {\ref{A3:update-omega1}}), we subtract the index set $\Omega_3$ of the MTDs cancellated in the current iteration from the index set $\Omega_1$ of the MTDs to be decoded in the following iterations, which is updated as $\Omega_1=\{\Omega_1\setminus \Omega_3\}$. Also, we obtain the index set of the MTDs that are already subtracted in the previous iterations, which is denoted as $\Lambda=\{\Omega_0\setminus \Omega_1\}$. Thirdly (line \ref{A3:update-H}), we update the measurement matrix as ${\bf H}={{\bf H}_0}_{[:,\{[KN_t]\setminus\widetilde{\Lambda}\}]}$, where ${\bf H}_0\in\mathbb{C}^{N_r \times (K N_t)}$ is the input channel matrix, $\widetilde{\Lambda}$ denotes the index set of the MAPs of the MTDs whose indices are in $\Lambda$, and the relationship between $\widetilde{\Lambda}$ and $\Lambda$ is defined in line \ref{A3:initial} of {\bf Algorithm \ref{Algorithm:3}}. Finally (lines \ref{A3:NoneDecision}$-$\ref{A3:endif}), if $\Omega_1$ is empty, the algorithm is terminated; otherwise, the updated ${\bf Y}$, ${\bf H}$, and $\Omega_1$ are fed back to line \ref{A3:JSAMP} to continue the next IDS-AMP iteration. Since the sparsity level in the next iteration is reduced with the unaltered dimension $N_r$ of the observations (i.e., the number of antennas at the BS), the proposed IDS-AMP detector is capable of achieving the improved decoding performance by using the aforementioned interference cancellation module.

\section{Channel Estimation for Media Modulation Based mMTC}
In this section, we first present the Gauss-Markov block fading channel model. Then, the CE is discussed in two stages. In the first stage, we discuss the initial CE stage to acquire the CSI of all the MTDs based on preambles. In the following stage, we propose a data-aided CSI update strategy to reduce the training overhead.

\begin{figure*}[t]
\vspace{-7mm}
\captionsetup{font={footnotesize, color = {black}}, justification = raggedright}
\centering
\includegraphics[scale=0.69]{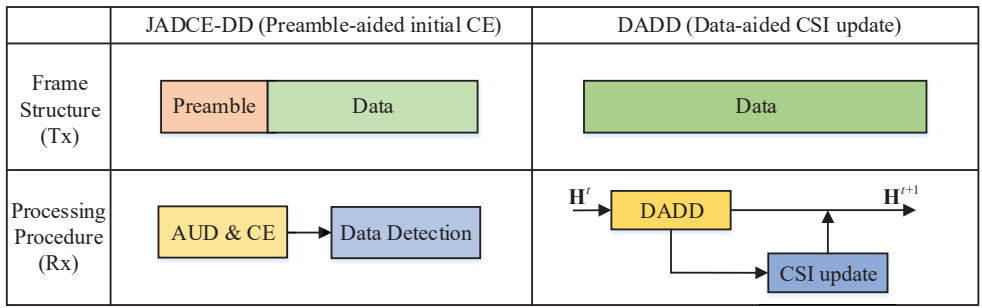}
\captionsetup{font={footnotesize, color = {black}}, justification = raggedright,labelsep=period}
\caption{Schematic diagram of the two-stage massive access scheme: the JADCE-DD stage and the following DADD stage, where the frame structure at the transmitter and the processing procedure at the receiver for each stage are presented in detail.}
\vspace{-4.5mm}
\label{Fig:CEsche}
\end{figure*}

Firstly, the Gauss-Markov block fading channel of the $k$-th MTD in the $(t+1)$-th frame (block), $\forall k\in[K]$, can be expressed as
\vspace{-2mm}
\begin{equation}\label{eq:channel_model}
\begin{split}
{\bf H}_k^{t+1}=\sqrt\alpha{\bf H}_k^{t}+\sqrt{1-\alpha}{\bf V}_k^{t},
\end{split}
\end{equation}
where ${\bf{H}}_k^t\in\mathbb{C}^{N_r\times N_t}$ is the MIMO channel matrix corresponding to the $k$-th MTD in the $t$-th time slot, elements in ${\bf{H}}_k^t$ obey the i.i.d. complex Gaussian distribution, and ${\bf V}_k^{t} \in \mathbb{C}^{N_r \times N_t}$ is the uncorrelated channel aging error. The elements in ${\bf V}_k^{t}$ also follow the i.i.d. complex Gaussian distribution with zero mean and unit variance \cite{BlockFading-ACCESS}. Furthermore, $\alpha$ is the autoregressive (AR) coefficient and the time-domain correlation is $\alpha^{\tau/2}\!\!\!\!=\!\!\!\!\frac{\mathbb{E}\left[[{\bf H}_k^{t}]_{n,i}^*[{\bf H}_k^{t+\tau}]_{n,i}\right]}{\mathbb{E}\left[[{\bf H}_k^{t}]_{n,i}^*[{\bf H}_k^{t}]_{n,i}\right]}$, where $\mathbb{E}\left[ \cdot\right]$ is the expectation operation, $n\!\in\![N_r]$, $i\!\in\![N_t]$ \cite{BlockFading-JSTSP}. The discrete time-lag $\tau\!\in \!\mathbb{Z}$ is equivalent to the number of symbols within a frame (block) \cite{BlockFading-JSTSP} and a case study will be presented in the simulations.

As for the initial CE at the first stage, the frame structure of the access signals consists a unique preamble allocated to each MTD and the following payload data, as shown in Fig. \ref{Fig:CEsche}. The active MTDs transmit their preambles and data to the BS without any grant. At the BS, joint activity detection and channel estimation (JADCE) as well as data detection (DD) is conducted in each frame. Due to the sporadic traffic, the JADCE-DD process can be realized by using CS techniques. The details of CS-based algorithms for JADCE-DD applied to media modulation aided grant-free massive access can be found in e.g., Section VI of \cite{MBMMUD4}. In particular, the authors of \cite{MBMMUD4} employed greedy CS algorithms, while other kinds of CS algorithms can also be utilized, e.g., the AMP-based algorithms according to \cite{KeMaLongTSP}, which is beyond the scope of this paper. Owing to the slowly time-varying CSI and the sporadic device activity \cite{JSAC-Editor}, the BS can obtain the CSI of all the MTDs after several frames.

At the following stage, as indicated in Fig. \ref{Fig:CEsche}, only data symbols are transmitted in successive frames and DADD is performed at the BS. Hence, we refer to this stage as the DADD stage. Note that the CSI utilized in the first frame of the DADD stage is obtained at the JADCE-DD stage. Furthermore, to combat the channel aging and to reduce the training overhead, we propose a CSI update strategy with the aid of the detected data symbols, as shown in Fig. \ref{Fig:CEsche}. Specifically, by providing ${\bf H}^t=[{\bf H} _1^t, {\bf H} _2^t,...,{\bf H} _K^t] \in \mathbb{C}^{N_r \times (K N_t)}$ and the received signal ${\bf{Y}}^t=\left[{{\bf {y}}_{1}^t, ..., {\bf {y}}_{J}^t}\right]\in\mathbb{C}^{N_r \times J}$ as the input of the proposed {\bf Algorithm 1} or {\bf Algorithm 3}, we can obtain the set of correctly detected active MTDs $\Omega^t$ and the reconstructed signal $\widehat{\bf X}^t\in \mathbb{C}^{KN_t\times J}$ of the $t$-th frame, $\forall t$. The number and the set of MAPs for the correctly estimated active MTDs are denoted as $N_a=|\Omega^t|_c$ ($N_a\le K_a$) and $\widetilde{\Omega}^t=\{\widetilde{\Omega}_u^t \}_{u=1}^{N_a}$, respectively, where $\widetilde{\Omega}_u^t=\{ N_t(\Omega^t|_u-1)+i\}_{i=1}^{N_t}$ is the set of MAPs for the $u$-th MTD in $\Omega^t$ for $u\in [N_a]$. Hence, we can transform the system model in (\ref{eq:systemModel}) as 
\begin{equation}\label{eq:feedback}
\begin{split}
{\bf Y}^t  \approx \widetilde{\bf H}^t\widetilde{\bf X}^t+{\bf W}^t,
\end{split}
\end{equation}
where $\widetilde{\bf X}^t \!=\! \widehat{\bf X}^t_{[\widetilde{\Omega}^t,:]}\!\in\!\! \mathbb{C}^{N_aN_t\times J}$ and $\widetilde{\bf H}^t\!=\!{\bf H}^t_{[:,\widetilde{\Omega}^t]}\!\in\!\! \mathbb{C}^{N_r\times N_aN_t}$ are the estimated media modulated signals and the associated CSI for active MTDs, respectively, and the approximately equal sign ``$\approx$'' can be written as an equal sign ``$=$'' if perfect DADD is achieved. Due to the slowly time-varying IoT channels, the frame length $J$ can be longer than $N_aN_t$, where $N_a\!\!\le\!\! K_a$ and $K_a$ is only a fraction of the total MTDs. Hence, we can refine the CSI $\widetilde{\bf H}^t$ based on $\widetilde{\bf X}^t$. For example, by using the MMSE estimator\footnote{Denote $P_{u}$ as the probability when the estimated media modulated signal matrix $\widehat{\bf X}^t_{[\widetilde{\Omega}_u^t,:]}\in \mathbb{C}^{N_t\times J}$ of given MTD, $\forall u\in [N_a]$, is a singular matrix. Then, considering the typical case, i.e., one row of $\widehat{\bf X}^t_{[\widetilde{\Omega}_u^t,:]}$ is all zeros, we have $P_{u}\approx JN_t(1-\frac{1}{N_t})^J$. If $J=200$ and $N_t=4$, we have $P_{u}\approx 8.2\times 10^{-23}$. Hence, it is suitable to assume that $\widetilde{\bf X}^t$ is a non-singular matrix, which enables us to adopt the MMSE estimator.}, 
we can obtain the refined CSI of active MTDs, denoted as $\widehat{\bf H}^t={\bf Y}^t \left( (\widetilde{\bf X}^t)^H {\bf R}_{\bf H}^t \widetilde{\bf X}^t + N_aN_t\sigma_w^2{\bf I}\right)^{-1}(\widetilde{\bf X}^t)^H{\bf R}_{\bf H}^t\in \mathbb{C}^{N_r\times N_aN_t}$, where ${\bf R}_{\bf H}^t=\mathbb{E}\left[(\widetilde{\bf H}^t)^H\widetilde{\bf H}^t\right]$ is the channel’s covariance matrix \cite{SM-CE-Hanzo}. Then, in the $(t+1)$-th frame, we set ${\bf H}^{t+1}={\bf H}^t\in \mathbb{C}^{N_r \times (K N_t)}$ and update part of it by performing ${\bf H}^{t+1}_{[:,\widetilde{\Omega}^t]}=\widehat{\bf H}^t$. Hence, we can adopt the updated CSI ${\bf H}^{t+1}$ as an input of the proposed {\bf Algorithm 1} or {\bf Algorithm 3} for DADD in the $(t+1)$-th frame.

\section{Performance Evaluation}
\subsection{Simulation Setup}
In this section, an extensive simulation investigation is carried out to evaluate the activity detection error rate (ADER) of the MTDs, the symbol error rate (SER), and the bit error rate (BER) of the proposed massive access solution. The ADER, SER, and BER are defined as
\begin{align}
{\rm ADER}&=\dfrac{E_m+E_f}{K},\\
{\rm SER}&=\dfrac{JE_m+E_{\rm symbol}}{J K_a},\\
{\rm BER}\!&=\!\dfrac{\eta J E_m +E_{\rm MED}+E_{\rm QAM}}{\eta J K_a },
\end{align}
respectively, where $E_m$ is the number of active MTDs missed to be detected, $E_f$ is the number of inactive MTDs falsely detected to be active, $E_{\rm symbol}$ is the number of error symbols\footnote{We consider that one symbol is in error if either its media modulated bits or quadrature amplitude modulated bits are in error.} of the detected active MTDs, $J K_a $ is the total number of symbols transmitted by the $K_a$ active MTDs within one frame, $E_{\rm MED}$ and $E_{\rm QAM}$ are the numbers of the errors of the media modulated bits and the quadrature amplitude modulated bits of the detected active MTDs within one frame, respectively, and $\eta J K_a$ is the number of total bits transmitted by the $K_a$ active MTDs within one frame. {\color{black}As for the CSI update strategy, given the $t$-th frame, $\forall t$, NMSE$_{\bf H}$ denotes the normalized mean-square errors (NMSE) between the updated CSI $\overline{\bf H}^t\in \mathbb{C}^{N_r \times (K N_t)}$ utilized for DADD and the true CSI ${\bf H}^t\in \mathbb{C}^{N_r \times (K N_t)}$. In addition, the NMSE between the reconstructed signal matrix $\widehat{\bf X}^t$ and the true signal matrix ${\bf X}^t$ is denoted as NMSE$_{\bf X}$. Specifically, NMSE$_{\bf H}$ and NMSE$_{\bf X}$ can be respectively expressed as

\begin{align}
{\rm NMSE}_{\bf H}&=\left\| \overline{\bf H}^t-{\bf H}^t \right\|_F\big{/}\left\|{\bf H}^t \right\|_F,\\
{\rm NMSE}_{\bf X}&=\left\| \widehat{\bf X}^t-{\bf X}^t \right\|_F\big/\left\|{\bf X}^t \right\|_F.
\end{align}
\vspace{-2mm}
}

In the simulations, the number of MTDs is $K=500$ with $K_a=50$ active MTDs, where each MTD adopts $N_{\rm RF}=2$ RF mirrors for media modulation and 4-QAM ($M=4$). Hence, the throughput is $\eta=N_{\rm RF}+{\rm log_2}M=4$ bpcu. Moreover, the number of receive antennas is $N_r=256$, the maximum iteration number is set to $T_0=15$.  Additionally, the frame length $J$ is set to 12 for uncoded media modulation based mMTC. For the SE of the DS-AMP algorithm, the number of Monte Carlo simulations is $N_{\rm MC}=500$, the maximum number of iterations is $T_{\rm SE}=50$, and the terminal threshold is $\varepsilon=10^{-5}$. Since we can obtain the {\it a posteriori} estimation of the media modulation signals ${\bf x}_{k,j}$, $\forall k,j$, in each Monte Carlo simulation (i.e., line \ref{Algo2:PostX} in {\bf Algorithm \ref{Algorithm:2}}), the ADER, BER, and SER of the theoretical SE can be calculated in the same way as those for the DS-AMP algorithm, and they can be then averaged over all Monte Carlo simulations. For the proposed coded media modulation based mMTC scheme, we consider a Turbo code with 1/3 rate and 12 tail bits. The length of the data packet is $L=120$ with the length of the signature sequence being $L_s=20$. \textcolor{black}{Hence, after channel encoding, the length of the data packet is $L'=3L+12=372$ and the frame length is $J=L'/ \eta=93$ for coded media modulation based mMTC.} In addition, $\overline{N}$ is set to 5 in line \ref{A3:OMEGA2} of {\bf Algorithm \ref{Algorithm:3}}. {\color{black}Without loss of generality, we investigate the ADER, SER, and BER performance of different algorithms in any given frame with perfect CSI at the BS, and a Rayleigh MIMO channel is considered.}

\subsection{Performance of the Proposed DS-AMP Algorithm}

For comparison, we consider the following benchmarks. {\bf Benchmark 1}: LMMSE multi-user detector for a traditional uplink mMIMO system \cite{Gao}, where $K_a$ single-antenna users (after the grant-based scheduling) adopting 16-QAM (for achieving the same throughput of 4 bpcu) are supported by an mMIMO BS with $N_r=256$ receive antennas. {\bf Benchmark 2}: The StrOMP algorithm (i.e., algorithm 1 in \cite{MBMMUD3}) is used for the activity detection and the SIC-SSP algorithm (i.e., algorithm 2 in \cite{MBMMUD3}) is used for data detection, where the terminal threshold $P_{\rm th}$ for the StrOMP algorithm is set to 1.5. {\bf Benchmark 3}: A modified DS-AMP algorithm without executing the min-max normalization (i.e, the min-max normalization in line \ref{A1:minmax} of {\bf Algorithm 1} is replaced by $\tilde{\bf a}=\widehat{\bf a}$), where the activity detection method is the same as that in \cite{AMP-AUD} (i.e., $\{k|[\widehat{\bf a}]_k>0.5\}$, $\forall k\in[K]$). {\bf AMP}: The conventional AMP algorithm \cite{AMPmeng} (i.e., only perform lines \ref{A1:initial}$-$\ref{A1:denoising} and \ref{A1:endfor} in {\bf Algorithm \ref{Algorithm:1}}), where the sparsity level is $\lambda=\frac{K_a}{K}$ and the noise variance $\sigma_w^2$ are perfectly known in advance, and the {\it a priori} probability in (\ref{eq:PriorMarginal}) is replaced by $p\left({\left[{{\bf x}_{k,j}}\right]_i}\right)=(1-\lambda)\delta\left({\left[{{\bf x}_{k,j}}\right]_i}\right)+\frac{\lambda}{M}\sum\nolimits_{s\in\mathbb{S}}\delta\left({\left[{{\bf x}_{k,j}}\right]_i-s}\right)$. {\bf TLSSCS}: The cutting-edge TLSSCS algorithm from \cite{TWOLEVEL}, where the scaling factor $\alpha=4$ (i.e., $\alpha$ in (6) of literature \cite{TWOLEVEL}). {\bf PIA-MSMP}: The state-of-the-art PIA-MSMP algorithm (i.e., algorithm 1 in \cite{MBMMUD4}) with the perfectly known sparsity level.

\begin{figure*}[t]
\vspace{-2mm}
\centering
\subfigure[]{
    \begin{minipage}[t]{0.33\linewidth}
        \centering
\label{fig:SNRPe}
        \includegraphics[width = 1\columnwidth,keepaspectratio]{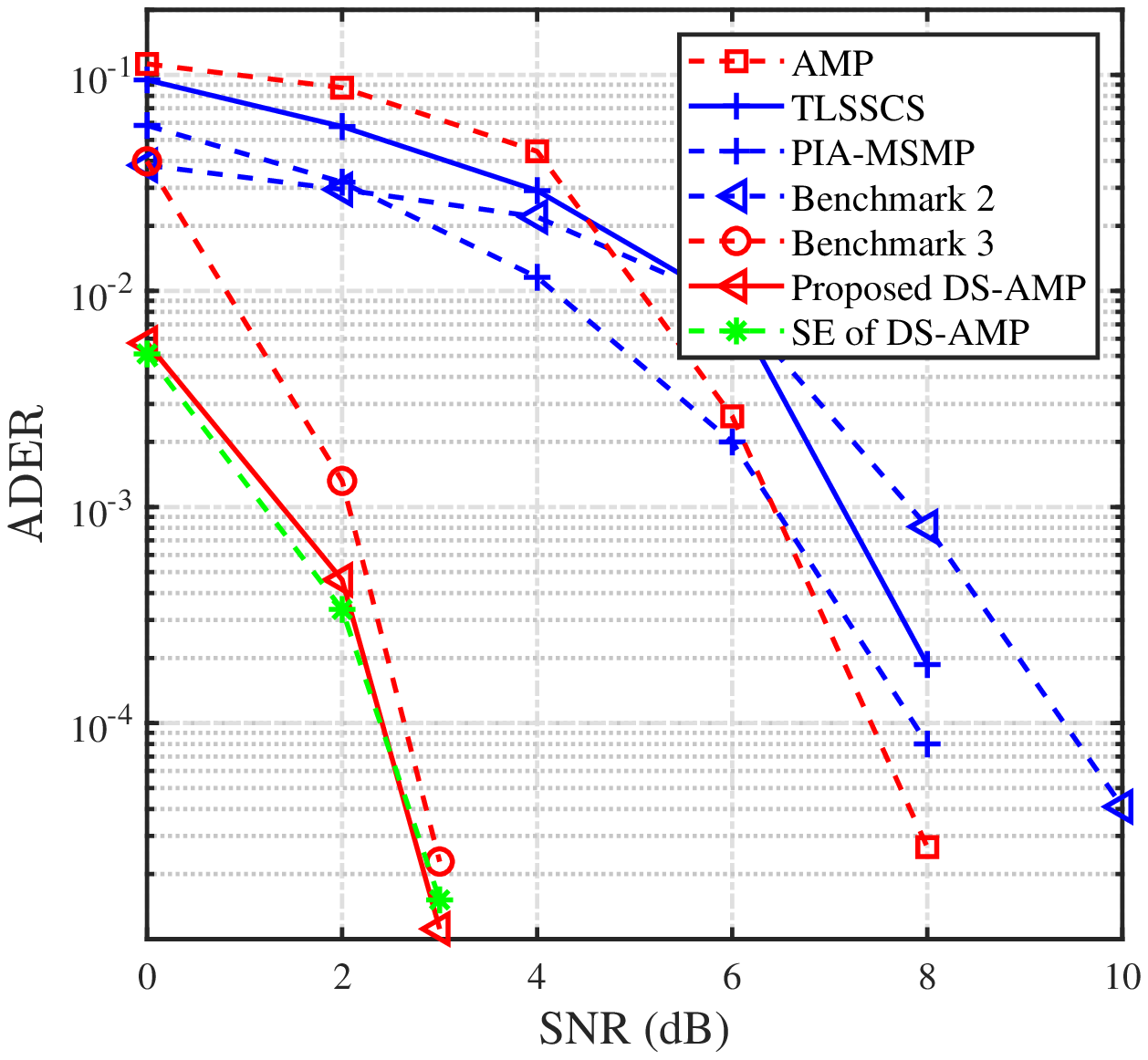}\\
    \end{minipage}%
}%
\subfigure[]{
    \begin{minipage}[t]{0.33\linewidth}
        \centering
\label{fig:SNRSER}
        \includegraphics[width = 1\columnwidth,keepaspectratio]{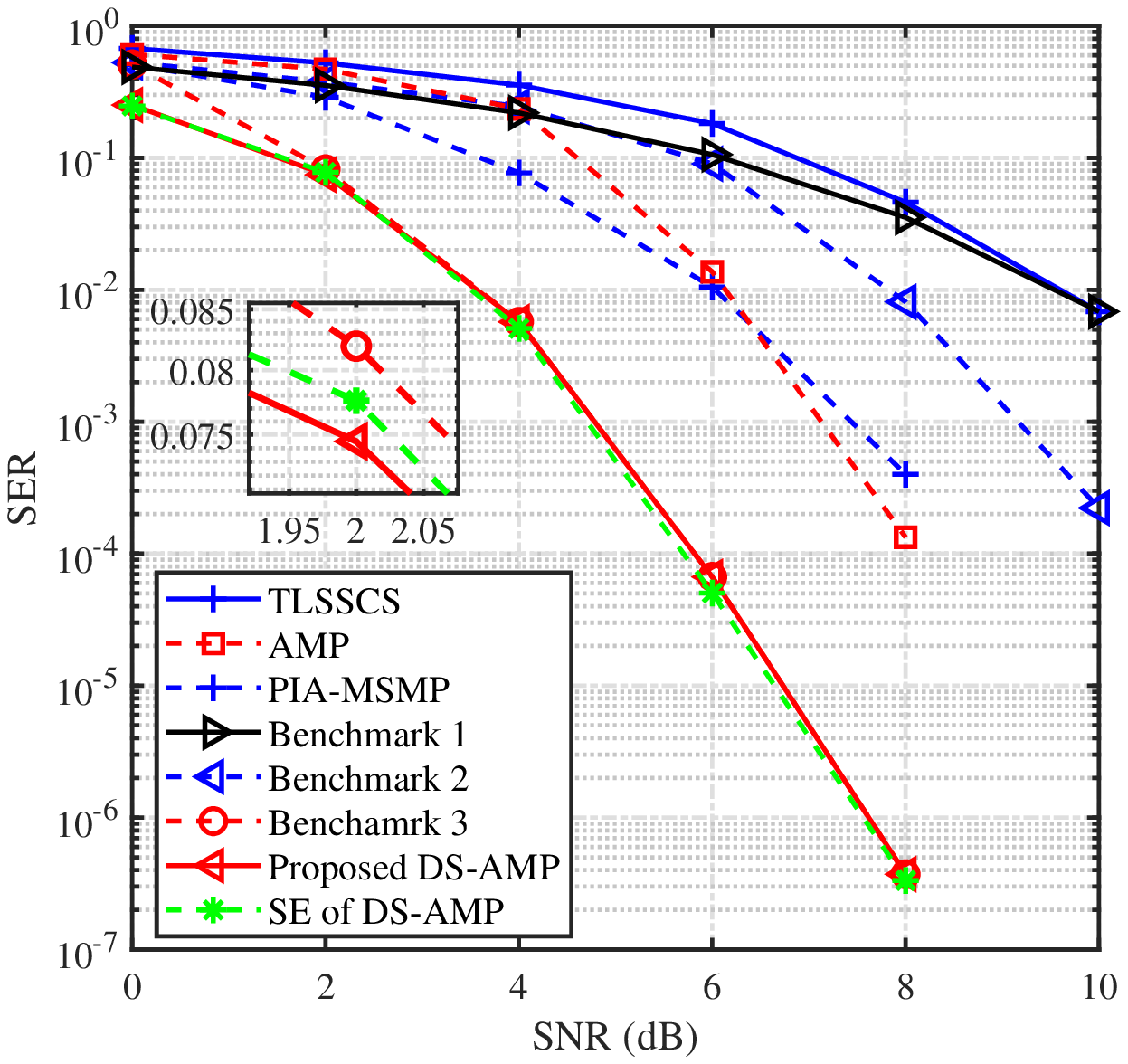}\\
    \end{minipage}%
}%
\subfigure[]{
    \begin{minipage}[t]{0.33\linewidth}
        \centering
\label{fig:SNRBER}
        \includegraphics[width = 1\columnwidth,keepaspectratio]{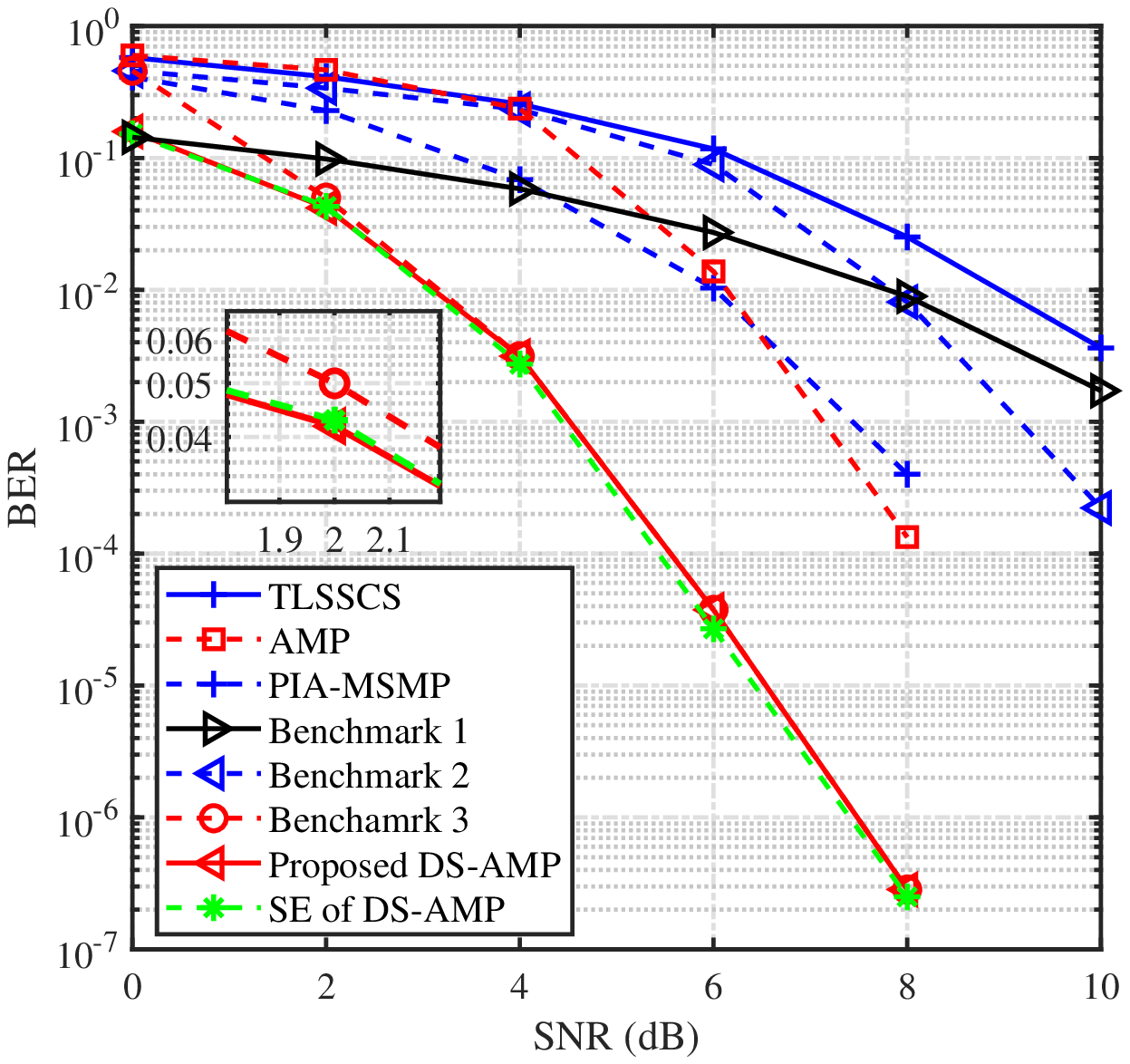}\\
    \end{minipage}%
}%
\centering
\setlength{\abovecaptionskip}{-1mm}
\captionsetup{font={footnotesize}, singlelinecheck = off, justification = raggedright,name={Fig.},labelsep=period}
\caption{Performance comparison of different solutions versus SNR: (a) ADER performance comparison; (b) SER performance comparison; (c) BER performance comparison. }
\label{fig:SNR}
\vspace{-5mm}
\end{figure*}

\begin{figure*}[t]
\centering
\subfigure[]{
    \begin{minipage}[t]{0.33\linewidth}
        \centering
\label{fig:TPe}
        \includegraphics[width = 1\columnwidth,keepaspectratio]{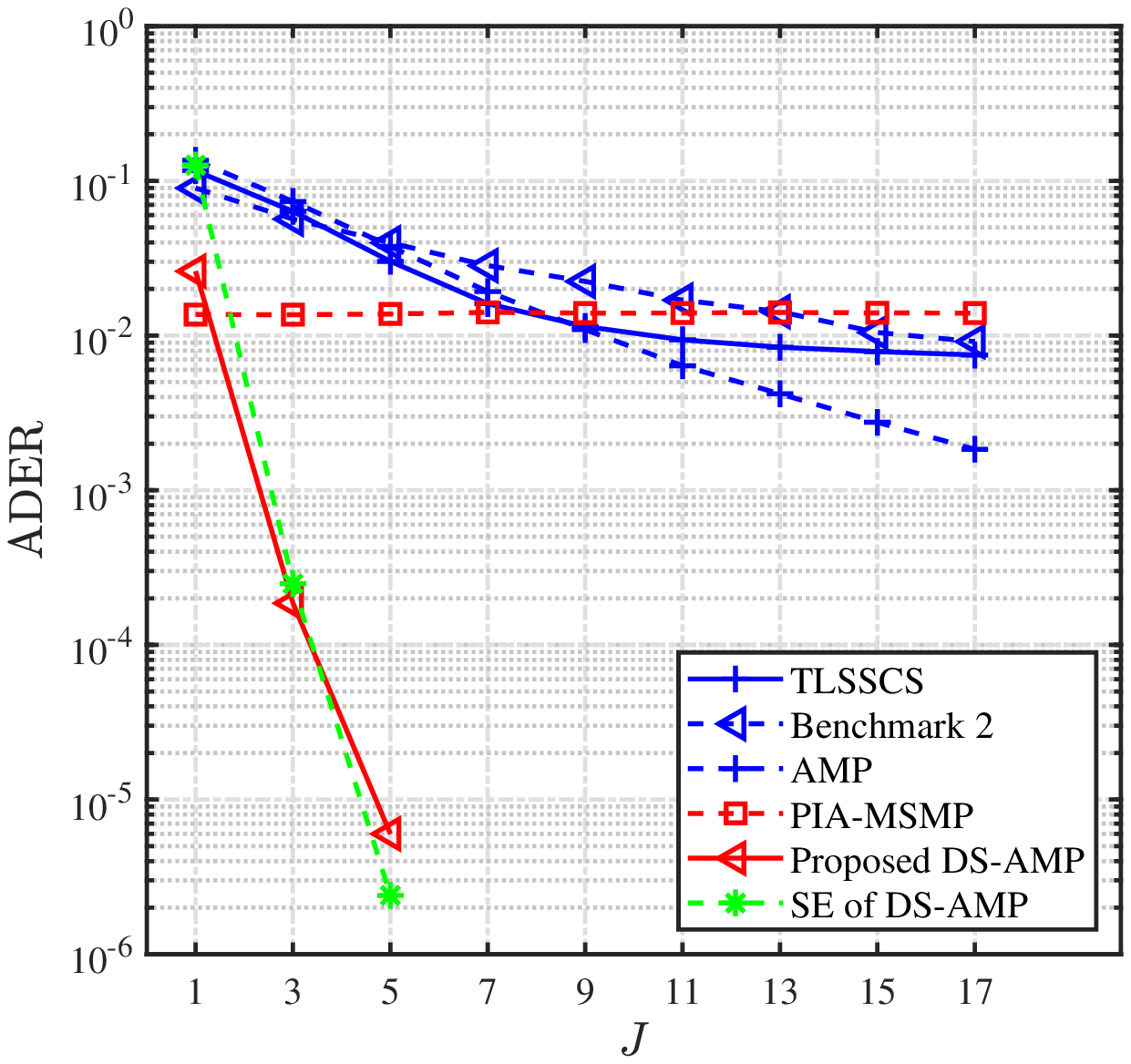}\\
    \end{minipage}%
                                                                                                                                                                                                                                                                                                                                                                                                                                                                                                                                                                                                                                                                                                                                                                                                                                                                                                                                                                                                                                                                                                                                                                                                                                                                                                                                                                                                                                                                                                                                                                                                                                                                                                                                                                                                                                                                        }%
\subfigure[]{
    \begin{minipage}[t]{0.33\linewidth}
        \centering
\label{fig:TSER}
        \includegraphics[width = 1\columnwidth,keepaspectratio]{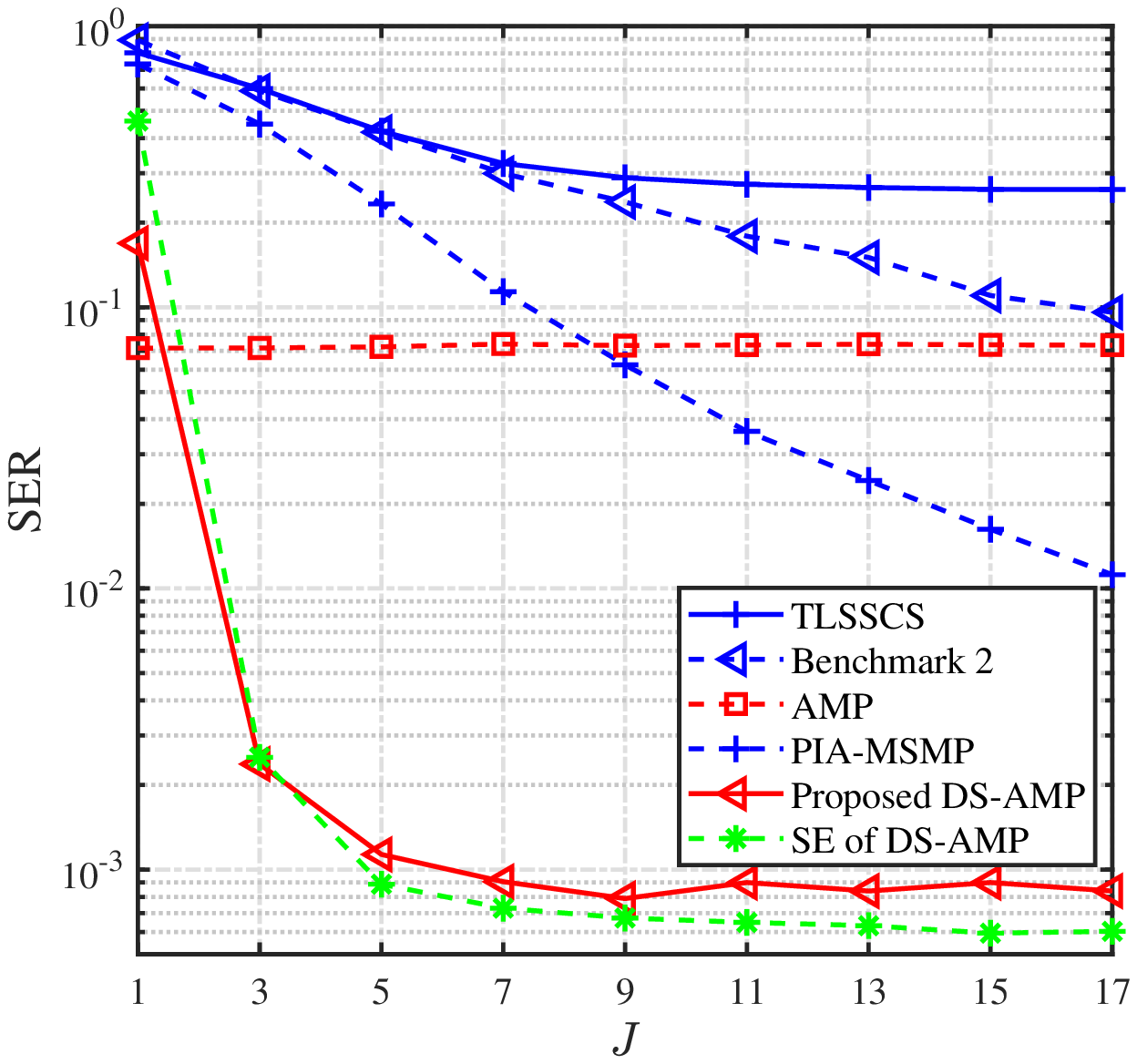}\\
    \end{minipage}%
}%
\subfigure[]{
    \begin{minipage}[t]{0.33\linewidth}
        \centering
\label{fig:TBER}
        \includegraphics[width = 1\columnwidth,keepaspectratio]{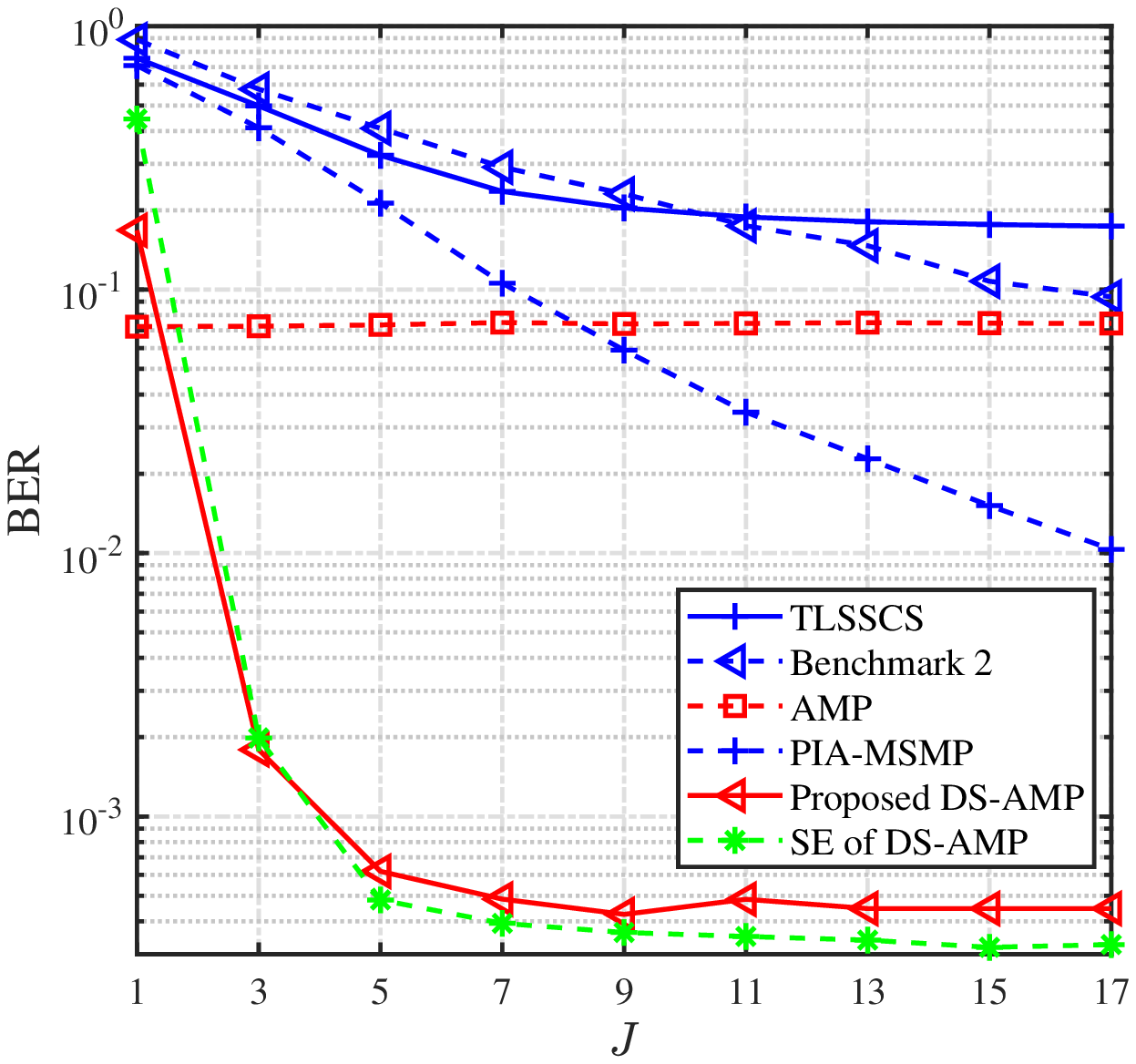}\\
    \end{minipage}%
}%
\centering
\setlength{\abovecaptionskip}{-1mm}
\captionsetup{font={footnotesize}, singlelinecheck = off, justification = raggedright,name={Fig.},labelsep=period}
\caption{Performance comparison of different solutions versus the numbers of time slots $J$ within a frame at SNR~$=$~5 dB: (a) ADER performance comparison; (b) SER performance comparison; (c) BER performance comparison. }
\label{fig:T}
\vspace{-6mm}
\end{figure*}

Fig. \ref{fig:SNRPe}, Fig. \ref{fig:SNRSER}, and Fig. \ref{fig:SNRBER} compare the ADER, SER, and BER performance of different solutions versus the SNR, respectively. Firstly, we observe that the SER and BER performance of the media modulation based mMTC scheme with the proposed DS-AMP algorithm outperforms the traditional uplink mMIMO scheme (benchmark 1) when the same throughput is considered. This verifies the advantages of the proposed media modulation based mMTC scheme over the traditional scheme. Since benchmark 1 perfectly knows the indices of the active MTDs, comparing the proposed scheme with benchmark 1 is actually unfair. Secondly, it can be observed that the proposed DS-AMP algorithm outperforms the TLSSCS algorithm, the PIA-MSMP algorithm, and benchmark 2 in terms of ADER, SER, and BER performance, which verifies the superiority of the proposed algorithm. Moreover, thanks to the exploitation of the doubly structured sparsity, our proposed DS-AMP algorithm outperforms the conventional AMP algorithm in ADER, SER, and BER. Furthermore, it is apparent that the proposed DS-AMP algorithm outperforms benchmark 3 in the low SNR regime (i.e., 0 dB$\sim$2 dB), which verifies the effectiveness of the proposed min-max normalization (i.e., line \ref{A1:minmax} in {\bf Algorithm \ref{Algorithm:1}}) in the DS-AMP algorithm. Finally, we observe that the SE offers a good tightness compared with the proposed DS-AMP algorithm in terms of ADER, SER, and BER performance, which can be observed from Fig. \ref{fig:T}, Fig. \ref{fig:S}, and Fig. \ref{fig:NR} as well.

From Fig. {\ref{fig:T}}, we observe that the proposed DS-AMP algorithm outperforms the conventional AMP algorithm, the TLSSCS algorithm, the PIA-MSMP algorithm, and benchmark 2 versus different frame lengths at SNR~$=$~5 dB in terms of ADER, SER, and BER performance. From Fig. {\ref{fig:TPe}}, owing to the exploitation of the structured sparsity in the time domain, it can be observed that the advantage of the proposed DS-AMP algorithm over other algorithms in terms of ADER performance becomes more evident upon increasing $J$. From Fig. {\ref{fig:TSER}} and Fig. {\ref{fig:TBER}}, if $J$ is small (i.e., $J<5$), we can obtain improved BER and SER performance as well as improved ADER performance. If $J$ is large (i.e., $J>9$), on the other hand, the BER and SER performance almost stays unaltered for different values of $J$. Since the conventional AMP algorithm reconstructs the signals of each time slot separately, its performance remains unchanged against different values of $J$.    

\begin{figure*}[t]
\vspace{-2mm}
\centering
\subfigure[]{
    \begin{minipage}[t]{0.33\linewidth}
        \centering
\label{fig:SPe}
        \includegraphics[width = 1\columnwidth,keepaspectratio]{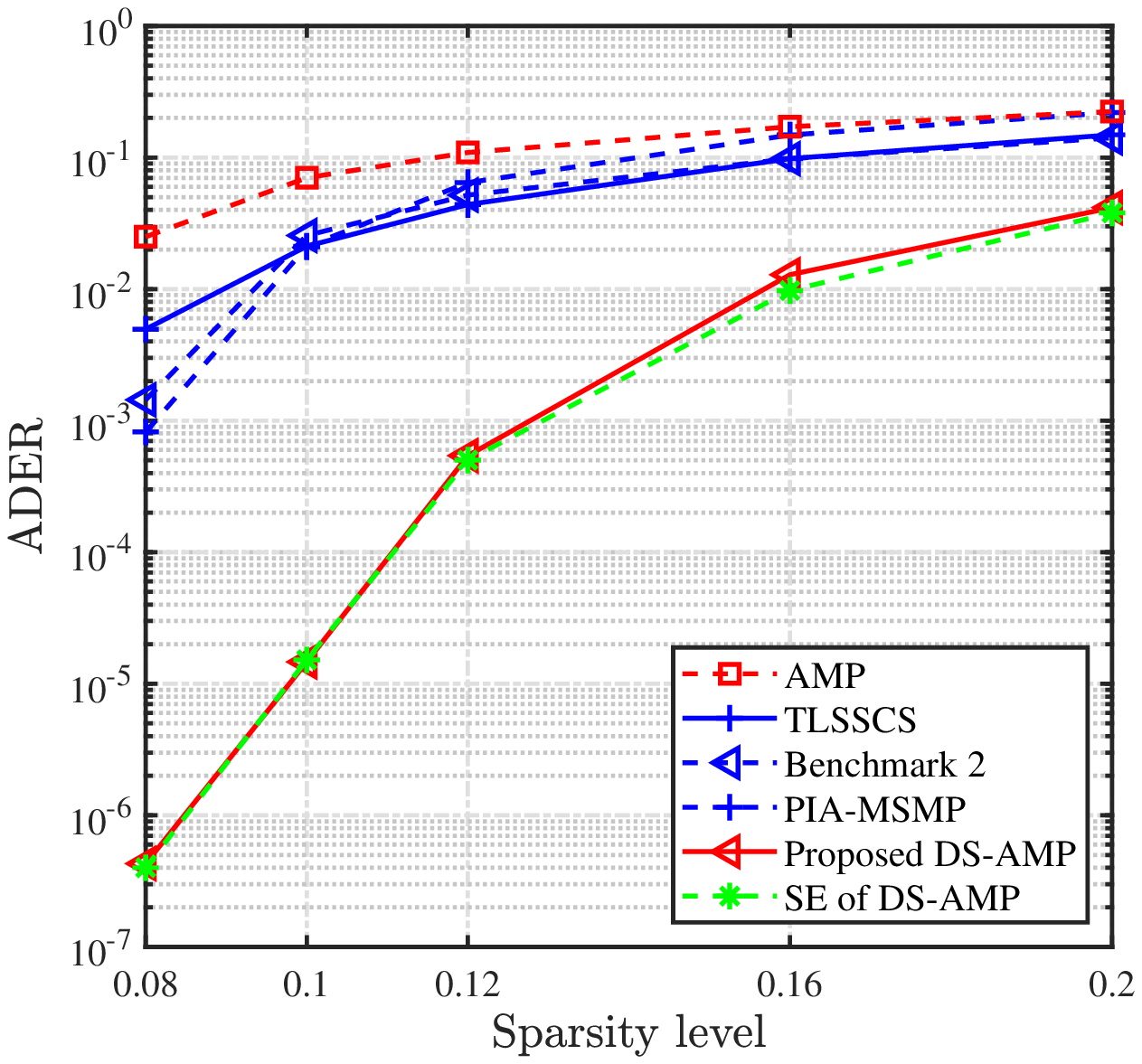}\\
    \end{minipage}%
}%
\subfigure[]{
    \begin{minipage}[t]{0.33\linewidth}
        \centering
\label{fig:SSER}
        \includegraphics[width = 1\columnwidth,keepaspectratio]{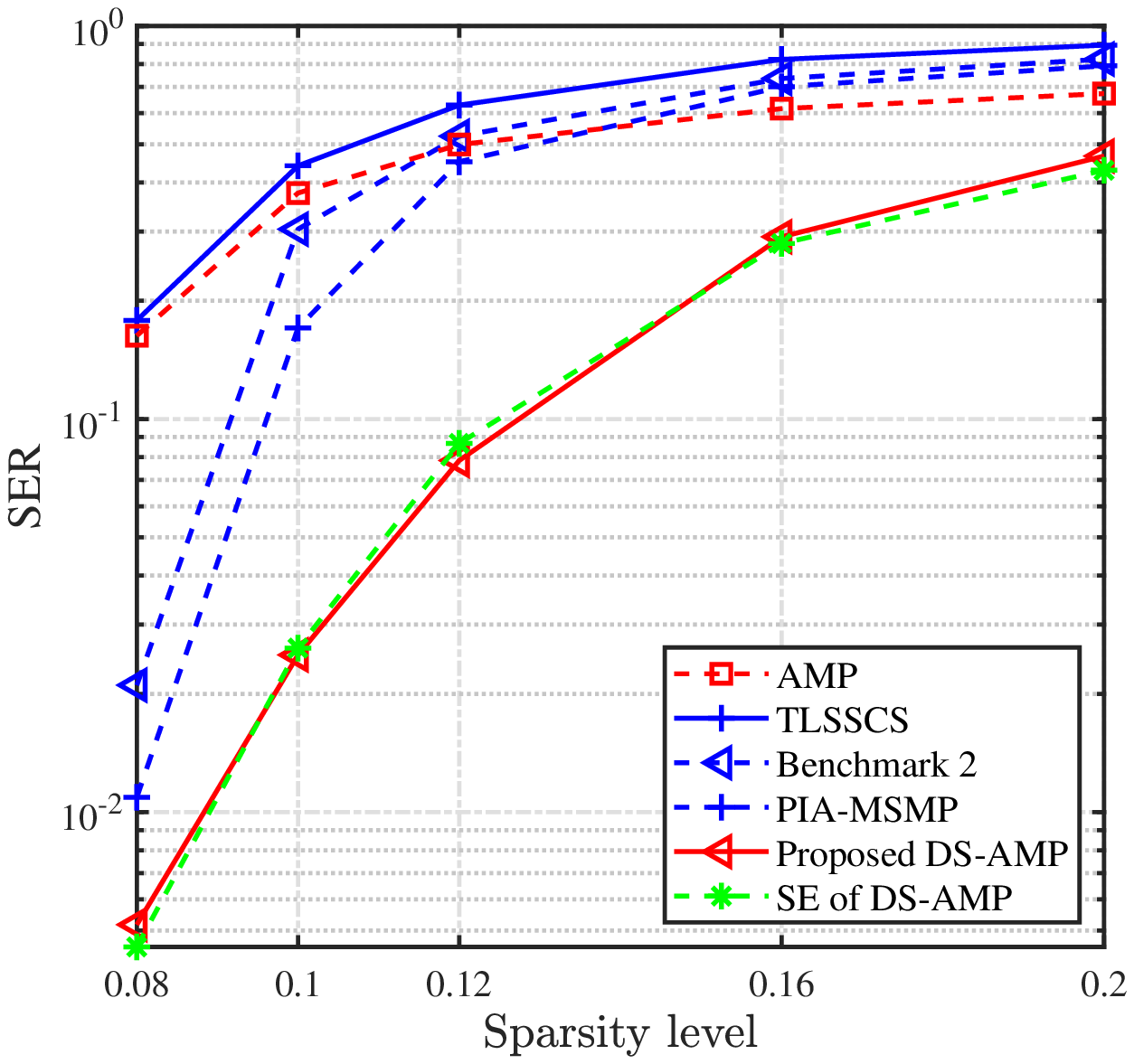}\\
    \end{minipage}%
}%
\subfigure[]{
    \begin{minipage}[t]{0.33\linewidth}
        \centering
\label{fig:SBER}
        \includegraphics[width = 1\columnwidth,keepaspectratio]{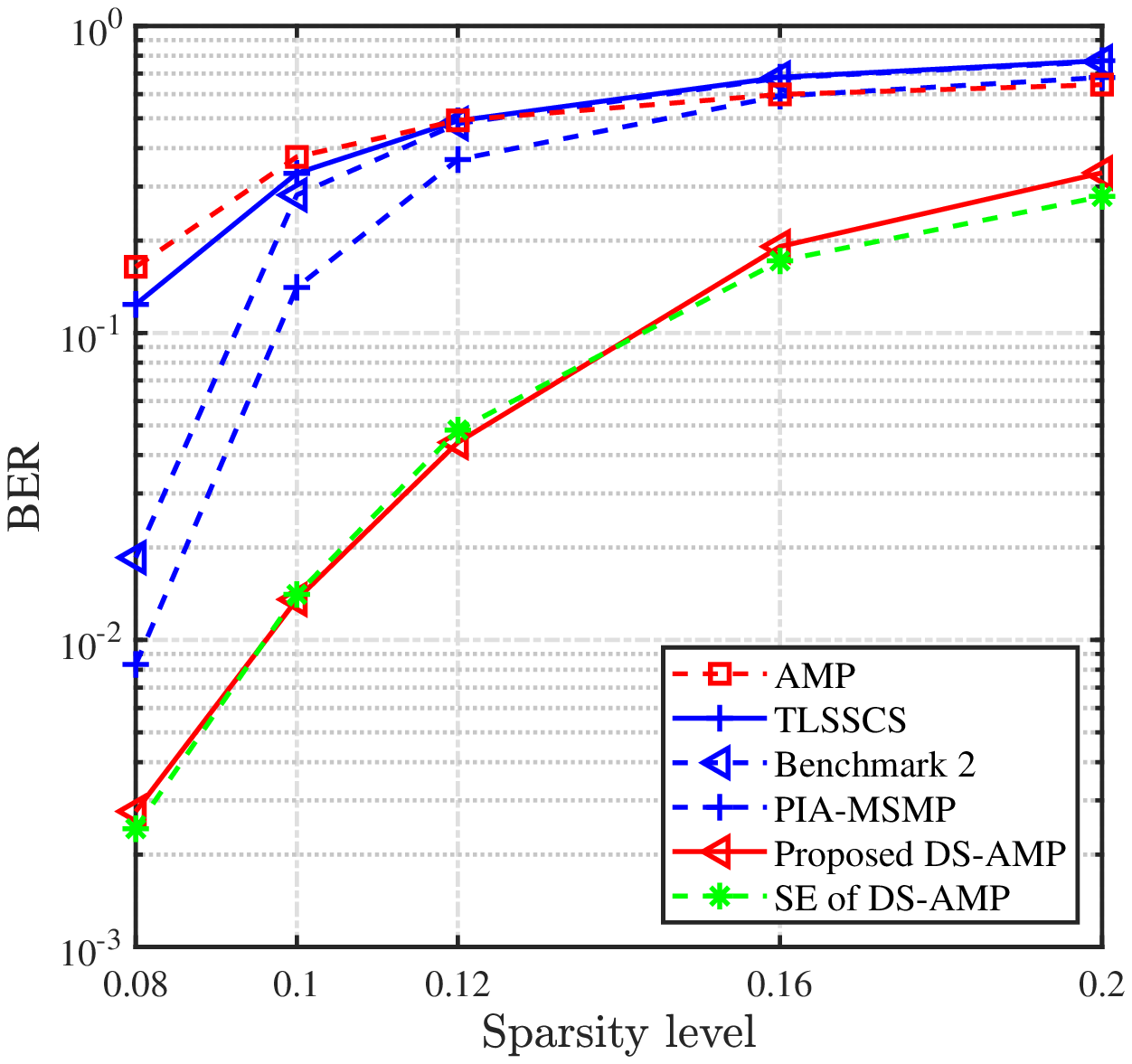}\\
    \end{minipage}%
}%
\centering
\setlength{\abovecaptionskip}{-1mm}
\captionsetup{font={footnotesize}, singlelinecheck = off, justification = raggedright,name={Fig.},labelsep=period}
\caption{Performance comparison of different solutions versus the sparsity level $\lambda=\frac{K_a}{K}$, given $K=500$ and SNR~$=$~3 dB: (a) ADER performance comparison; (b) SER performance comparison; (c) BER performance comparison. }
\label{fig:S}
\vspace{-2mm}
\end{figure*}

\begin{figure*}[t]
\vspace{-2mm}
\centering
\subfigure[]{
    \begin{minipage}[t]{0.33\linewidth}
        \centering
\label{fig:NRPe}
        \includegraphics[width = 1\columnwidth,keepaspectratio]{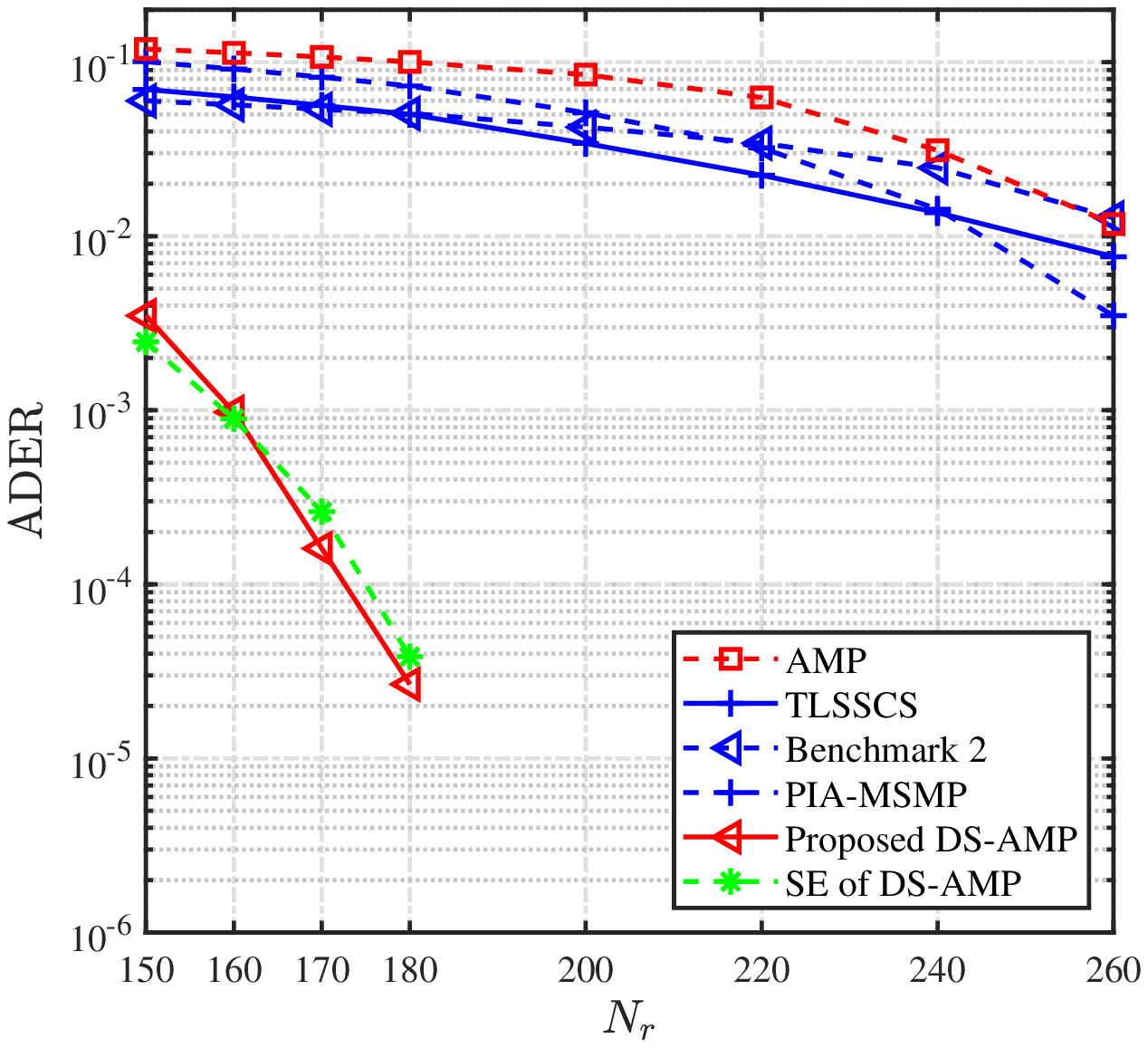}\\
    \end{minipage}%
}%
\subfigure[]{
    \begin{minipage}[t]{0.33\linewidth}
        \centering
\label{fig:NRSER}
        \includegraphics[width = 1\columnwidth,keepaspectratio]{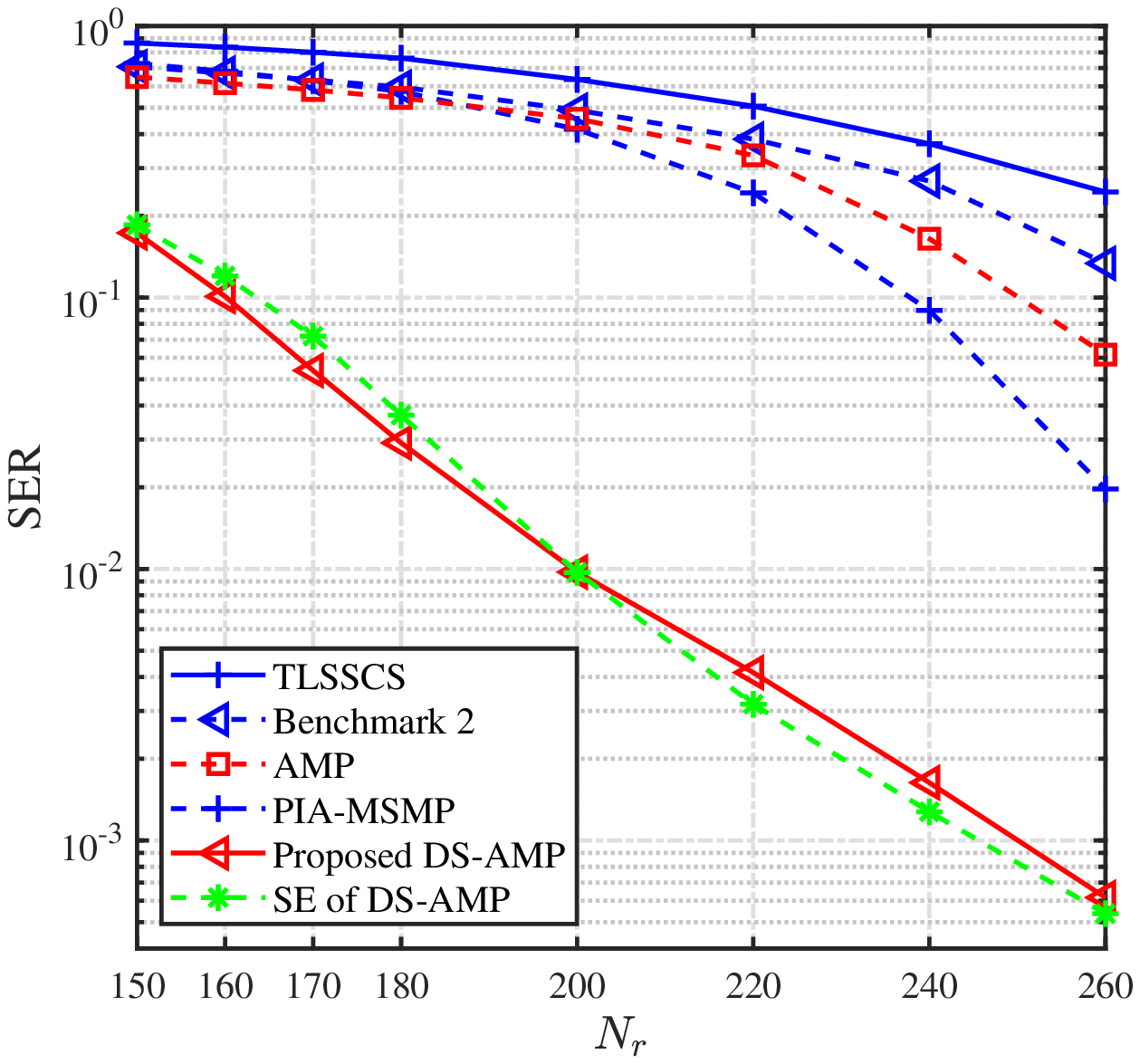}\\
    \end{minipage}%
}%
\subfigure[]{
    \begin{minipage}[t]{0.33\linewidth}
        \centering
\label{fig:NRBER}
        \includegraphics[width = 1\columnwidth,keepaspectratio]{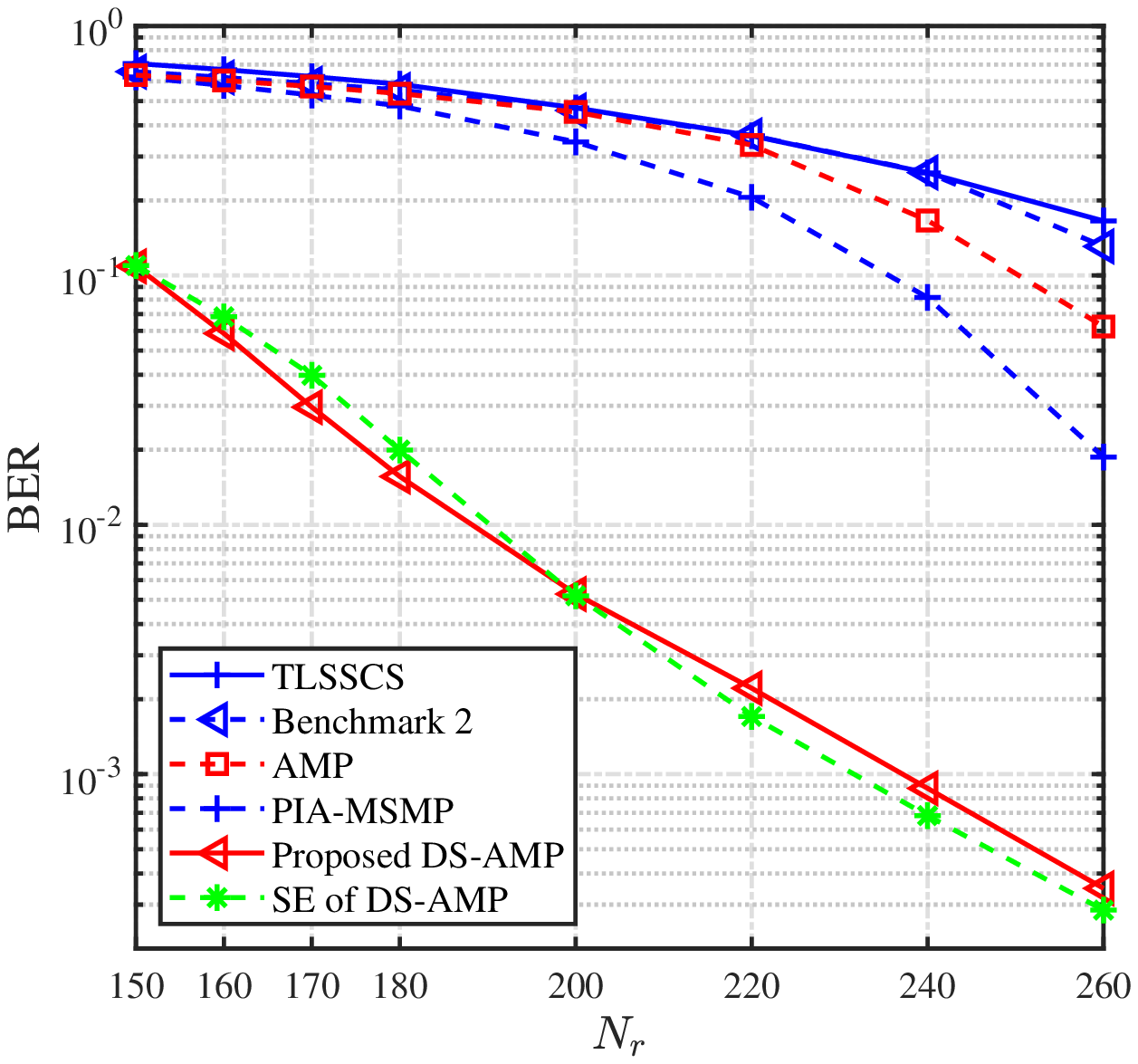}\\
    \end{minipage}%
}%
\centering
\setlength{\abovecaptionskip}{-1mm}
\captionsetup{font={footnotesize}, singlelinecheck = off, justification = raggedright,name={Fig.},labelsep=period}
\caption{Performance comparison of different solutions versus the numbers of receive antennas $N_r$ at SNR~$=$~5 dB: (a) ADER performance comparison; (b) SER performance comparison; (c) BER performance comparison. }
\label{fig:NR}
\vspace{-6mm}
\end{figure*}

Fig. \ref{fig:S} provides the ADER, SER, and BER performance comparisons of different algorithms versus the sparsity level ($\lambda=\frac{K_a}{K}$ given $K=500$ and SNR~$=$~3 dB). From Fig. \ref{fig:SPe}, Fig. \ref{fig:SSER}, and Fig. \ref{fig:SBER}, we observe that the proposed DS-AMP algorithm outperforms the conventional AMP algorithm, the TLSSCS algorithm, the PIA-MSMP algorithm, and benchmark 2 in terms of ADER, SER, and BER performance, respectively. Fig. \ref{fig:NR} depicts the ADER, SER, and BER performance of different algorithms versus the numbers of receive antennas $N_r$ at SNR~$=$~5 dB. From Fig. \ref{fig:NR}, similar conclusion as observed in Fig. \ref{fig:S} can be obtained. In particular, both figures verify the superiority and robustness of the proposed DS-AMP algorithm under different system parameters, i.e., the sparsity level or the number of receive antennas, in typical IoT scenarios.

Fig. 10 depicts the ADER and BER performance of the proposed DS-AMP algorithm versus the maximum iteration number $T_0$. We observe that the ADER and BER performance of the proposed DS-AMP algorithm converges fast at various SNRs (usually fewer than 15 iterations are needed), which indicates that we can adopt the maximum iteration number $T_0=15$ for {\bf Algorithm \ref{Algorithm:1}}. In particular, the SER performance of the DS-AMP algorithm versus $T_0$ is similar to that of the BER and ADER performance.

The computational complexity comparison of different solutions is provided in Table I. It is apparent that the computational complexity of the proposed DS-AMP algorithm is an order of magnitude lower than that of the TLSSCS algorithm, the PIA-MSMP algorithm, and benchmark 2 under the considered simulation parameters. Moreover, the complexity of the DS-AMP algorithm scales linearly with the number of receive antennas $N_r$, whereas the computational complexity of the TLSSCS and PIA-MSMP algorithms can be approximately proportional to the square of $N_r$. Hence, the proposed DS-AMP is more attractive than other state-of-the-art algorithms for solving the massive access problem in mMIMO systems. Furthermore, compared with the conventional AMP algorithm and benchmark 3, the proposed DS-AMP algorithm achieves better performance without substantially increasing the computational complexity.

\begin{table*}[!t]
\vspace{-8mm}
\scriptsize
\centering
\captionsetup{font = {normalsize, color = {black}}, labelsep = period} 
\caption*{Table I: Computational complexity comparison of different algorithms for uncoded media modulation based mMTC}
\begin{threeparttable}
\begin{tabular}{|p{2.8cm}|p{3cm}|p{9cm}|p{1.8cm}|p{1.8cm}|}
\Xhline{1.2pt}
\multicolumn{2}{|c|}{\multirow{2}*{{\bf Algorithms}}} & \multirow{2}*{{\bf Computational complexity}}& \multicolumn{2}{|c|}{{\bf Complex-valued multiplications\tnote{1}}}  \\%
\cline{4-5}
\multicolumn{2}{|c|}{~} & ~& $N_r=128$ &$N_r=256$  \\%
\Xhline{1.2pt}
\multicolumn{2}{|c|}{Benchmark 1} & $\mathcal{O}(JN_rK_a+2N_r{K_a}^2+{K_a}^3)$ &$0.84\times10^6$ & $1.56\times10^6$\\
\hline
\multicolumn{2}{|c|}{DS-AMP} &${\cal O}[T_0JKN_t(\frac{5}{2}N_r+|\mathbb{S}|_c+\frac{1}{4})]$ & $1.17\times10^8$ & $2.32\times10^8$\\
\hline
\multicolumn{2}{|c|}{AMP}&${\cal O}[T_0JKN_t(\frac{5}{2}N_r+|\mathbb{S}|_c+\frac{1}{4})]$ &$1.17\times10^8$ & $2.32\times10^8$\\
\hline
\multicolumn{2}{|c|}{Benchmark 3} &${\cal O}[T_0JKN_t(\frac{5}{2}N_r+|\mathbb{S}|_c+\frac{1}{4})]$  &$1.17\times10^8$ & $2.32\times10^8$\\
\hline
\multicolumn{2}{|c|}{TLSSCS} & $\mathcal{O}\{(JN_rK_a+2N_r{K_a}^2+{K_a}^3)+(K_a+1)[{N_r}^2(KN_t+J)+N_rJKN_t]+\sum\nolimits_{s=1}^{K_a+1}[{N_r}^2+2N_r(sN_t)^2+(sN_t)^3]\}$& $2.14\times10^9$ & $7.53\times10^9$\\
\hline
\multicolumn{2}{|c|}{PIA-MSMP} & $\mathcal{O}\{3JK_aN_r(N_t+1)+(K_a+1)[{N_r}^2(KN_t+J)+N_rJKN_t]+\sum\nolimits_{s=1}^{K_a}[{N_r}^2+2N_r(sN_t)^2+(sN_t)^3]\}$&$2.12\times10^9$ & $7.50\times10^9$\\
\hline
\multicolumn{2}{|c|}{Benchmark 2} &${\cal O}\{K_aJKN_tN_r+\sum\nolimits_{s=1}^{K_a}[JN_r(s+2s^2+2(sN_t)^2)+J(s^3+(sN_t)^3)]+\sum\nolimits_{s=1}^{K_a}[JN_r(s+2s^2+2(sN_t)^2)+J(s^3+(sN_t)^3)]\}$  &$4.82\times10^9$ & $8.16\times10^9$\\
\hline
\Xhline{1.2pt}
\end{tabular}
\begin{tablenotes}
\scriptsize
\item[1] The order of complex-valued multiplications is obtained under parameters $J=12$, $N_t=4$, $K=500$, $K_a=50$, $T_0=15$, $|\mathbb{S}|_c=4$.
\end{tablenotes}
\end{threeparttable}
\end{table*}

\begin{figure*}[t]
\vspace{-2mm}
\captionsetup{font={footnotesize, color = {black}}, justification = raggedright}
\centering
\begin{minipage}[c]{0.29\textwidth}
\centerline{\includegraphics[scale=.4]{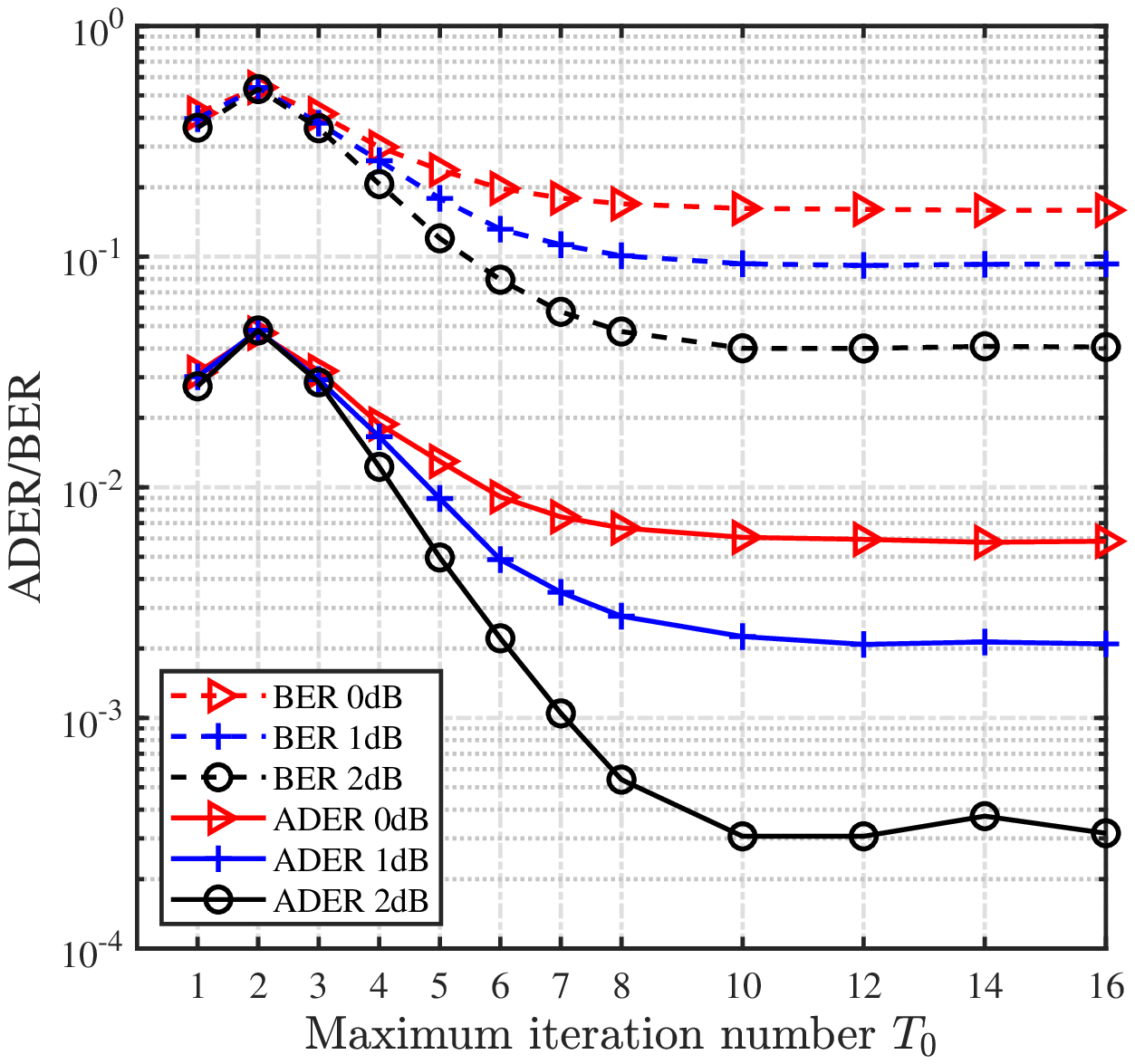}}
\caption*{Fig. 10. Performance of the proposed DS-AMP algorithm versus the maximum iteration number $T_0$.}
\label{fig:I}
\end{minipage}
\hfill
\begin{minipage}[c]{0.69\textwidth}
\centering%
\includegraphics[scale=.42]{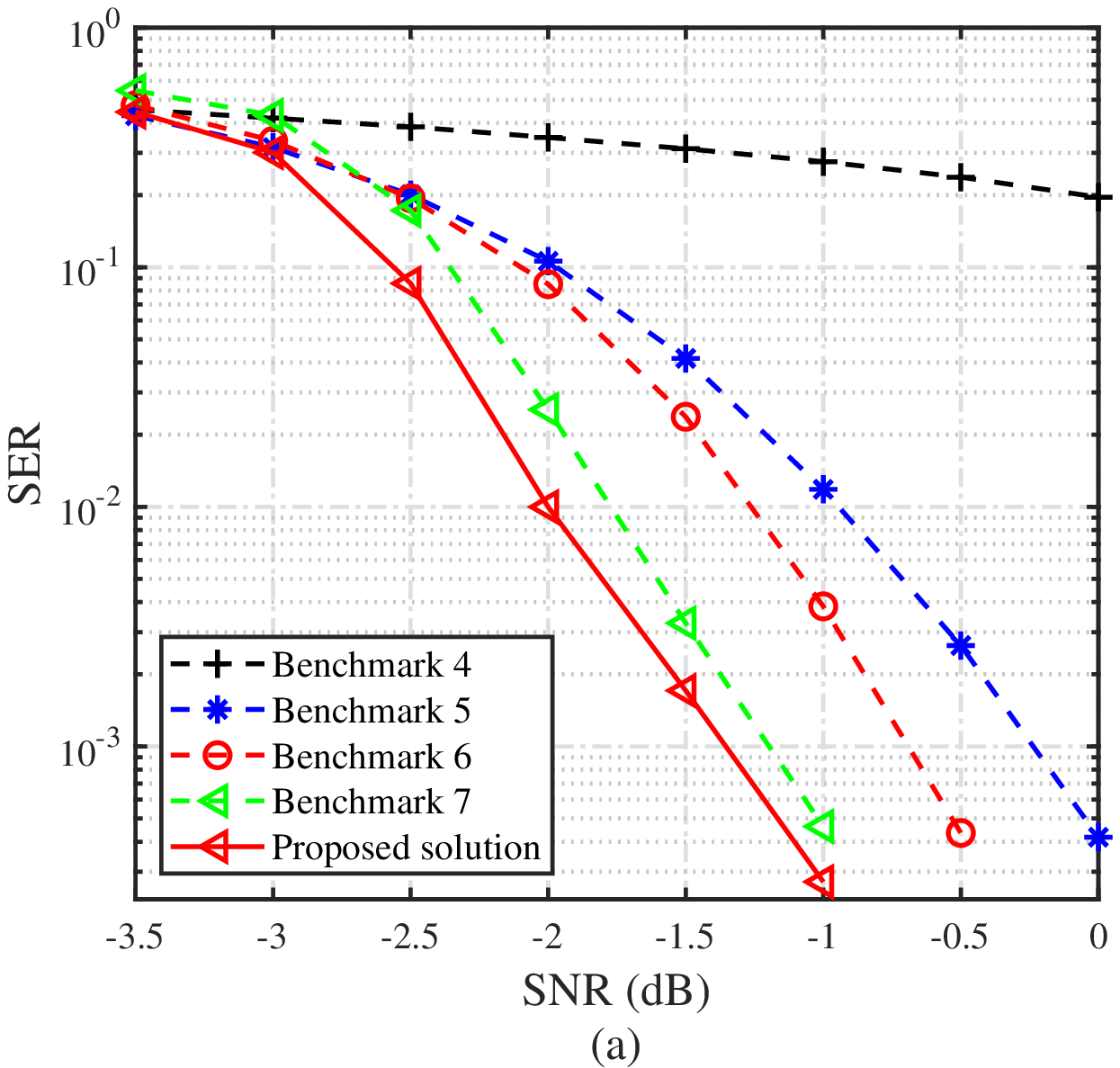}
\includegraphics[scale=.42]{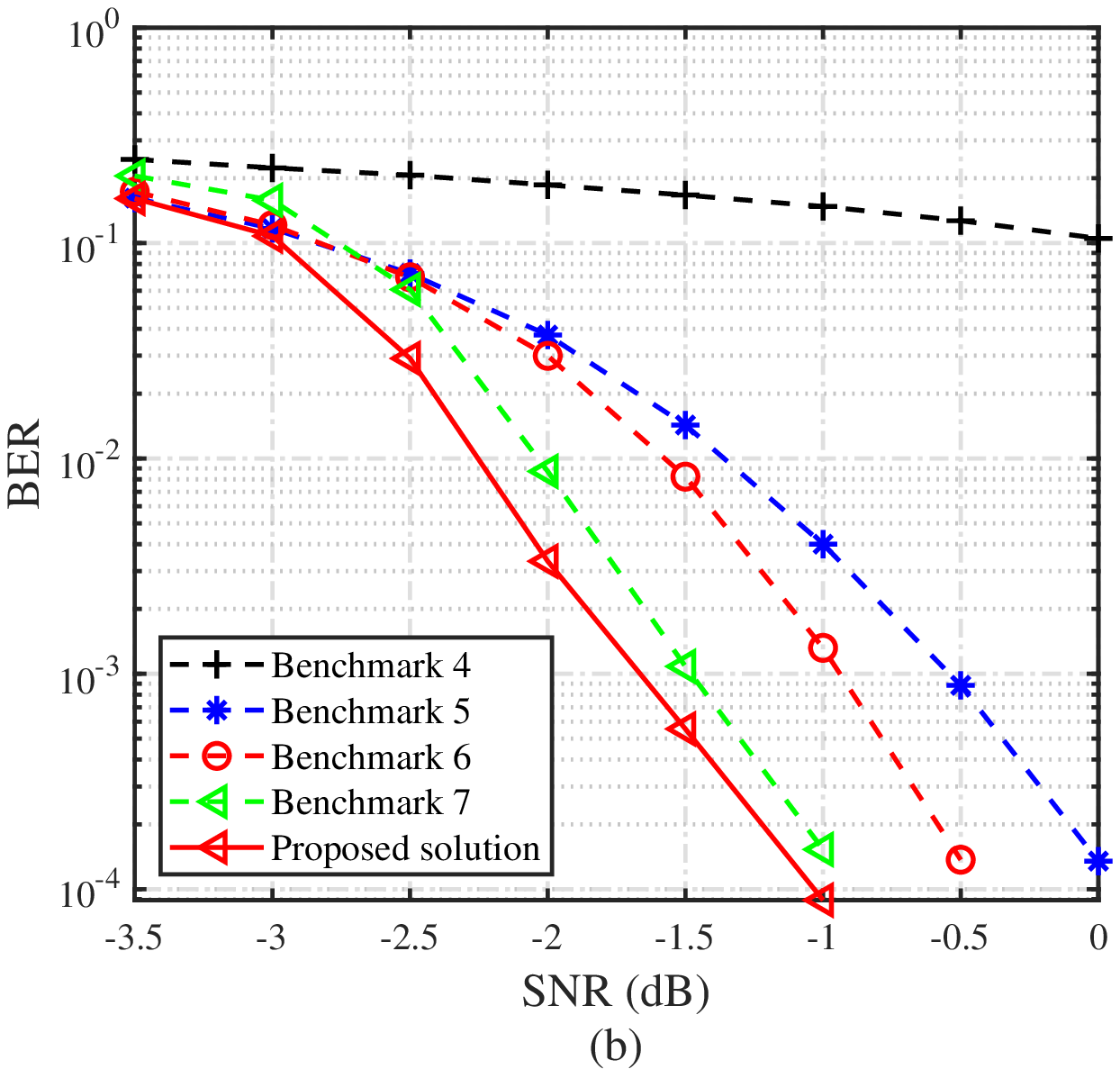}
\caption*{Fig. 11. Performance of the proposed SIC-based massive access solution in comparison with the benchmarks: (a) SER performance comparison; (b) BER performance comparison.}
\end{minipage}
\vspace{-6mm}
\end{figure*}

%
\vspace{-2mm}
\subsection{Performance of the Proposed IDS-AMP Scheme}

In this subsection, we compare the SER and BER performance of the proposed IDS-AMP scheme with the following benchmarks. {\bf Benchmark 4}: The proposed DS-AMP algorithm adopting uncoded media modulation and a hard decision (i.e., it performs a hard decision according to the output signal ${\bf X}\in\mathbb{C}^{KN_t\times J}$ from {\bf Algorithm 1} to get the demodulated binary bits). {\bf Benchmark 5}: The proposed DS-AMP algorithm adopting coded media modulation and soft decision, while the processing of interleaving/deinterleaving and SIC in Fig. \ref{fig:Communication} is not adopted. {\bf Benchmark 6}: The proposed DS-AMP algorithm adopting coded media modulation, interleaving/deinterleaving, and soft decision, while the SIC processing is not adopted. {\bf Benchmark 7}: The proposed IDS-AMP scheme with the exception of the proposed decoding quality judgement (i.e., lines \ref{A3:DQJ}$-${\ref{A3:WeightDecision} of {\bf Algorithm 3}) is removed and $\Omega_3$ in line \ref{A3:SICstart} is equal to $\Omega_2$.

Fig. 11 compares the SER and BER performance of benchmarks 4$-$7 and the proposed IDS-AMP scheme. The worst performance is achieved by benchmark 4, which indicates the necessity of adopting channel coding and soft decoding for improving the data decoding performance. The superiority of benchmark 6 over benchmark 5 verifies the efficiency of the proposed BICMM in overcoming the effect of burst error in spatial-selective channel fading. Besides, the superiority of benchmark 7 over benchmark 6 in the high SNR regime (i.e., greater than -2 dB) verifies the effectiveness of SIC processing in improving the data decoding performance. However, benchmark 7 is observed to be inferior to benchmark 5 in the low SNR regime (i.e., -3.5 dB$\sim$-2.5 dB), since the SIC at a low SNR can degrade the performance. By contrast, the proposed IDS-AMP scheme shows a consistent superiority over the four considered benchmarks. Particularly, the superiority of the IDS-AMP scheme over benchmark 7 verifies the data decoding gain achieved by the proposed decoding quality judgement module.

\vspace{-3mm}
\subsection{Performance of the CSI Update Strategy}
To investigate the data-aided CSI update strategy, we consider successive $N_f$ ($N_f\gg 1$) frames (blocks), where the Gauss-Markov block fading channel model (\ref{eq:channel_model}) is considered and a Rayleigh MIMO channel model is utilized for the first frame.
The coherence time $T_c$ is defined by the duration that the time-domain correlation function is above 0.5 \cite{BlockFading-JSTSP} and we have $T_c=0.423{/}f_m$, where $f_m={vf_c}{/}{v_c}$ is the maximum Doppler shift, $v$ is the maximum velocity of the MTDs, and $v_c$ is the speed of light \cite{BlockFading-JSTSP}. According to 3GPP  Narrowband Internet-of-Things (NB-IoT) specifications, $W=15$ kHz  bandwidth for single-tone uplink transmission with carrier frequency $f_c=900$ MHz can be supported \cite{NB-IOT}. In this case, the time-lag $\tau=\lfloor{T_c/T_s}\rfloor=213$, if $v=35.6$ km/h, where $T_s=1/W$ denotes the symbol duration. Since the time-domain correlation $\alpha^{\tau/2}$ has to be above 0.5 between the time indices $i=1$ and $i=\tau+1$, we obtain the AR coefficient  $\alpha=0.9935$ when $\alpha^{\tau/2}$ is set to 0.5 \cite{BlockFading-JSTSP}. Furthermore, the frame length after channel coding is $J=213$, and the associated number of bits transmitted by a frame is 280 bits. The AR coefficient is 0.99. The encoder and other system parameters are the same as that in Section VI-A, and the proposed DS-AMP algorithm with channel coding is used for DADD in each frame. 

\begin{figure}[t]
\captionsetup{font={footnotesize, color = {black}}, justification = raggedright}
\centering
\includegraphics[scale=0.8]{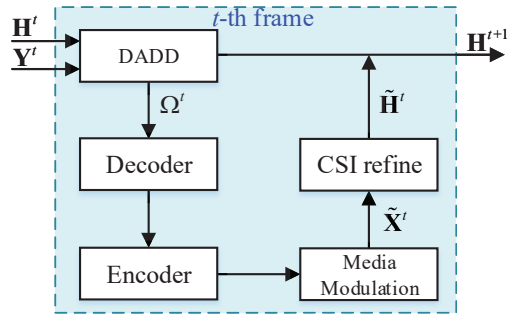}
\captionsetup{font={footnotesize, color = {black}}, justification = raggedright,labelsep=period}
\caption*{Fig. 12. The CSI update process of the $t$-th frame, $0<t<N_f$.}
\vspace{-8mm}
\end{figure}

For comparison, we consider two strategies, both of which assume that the CSI of all the MTDs has been acquired at the BS in the first frame. In particular, for the {\bf CSI non-update strategy}:  the CSI used for DADD in each frame is the same as that in the first frame. As for the {\bf proposed CSI update strategy}: the CSI of correctly detected active MTDs are refined by the MMSE estimator at the BS in the $t$-th frame, $0<t<N_f$, as shown in (\ref{eq:feedback}), and then the updated CSI matrix of all the MTDs is used for DADD in the $(t+1)$ frame. In particular, by decoding, re-encoding, and media modulating the bits associated with the detected active MTDs, as shown in Fig. 12, we can obtain the signal matrix $\widetilde{\bf X}^t$ for refining the CSI matrix $\widetilde{\bf H}^t$. Then, ${\bf H}^{t+1}$ is updated as shown in (\ref{eq:feedback}).

It can be seen from Fig. 13, as the time increases, the NMSE$_{\bf H}$ and NMSE$_{\bf X}$ performance of the CSI non-update strategy decreases monotonically due to the channel aging. On the contrary, by exploiting the proposed CSI update strategy, the NMSE$_{\bf H}$ and NMSE$_{\bf X}$ performance dramatically outperforms that of the non-update strategy, especially, after a relatively large number of frames. Hence, by using the proposed CSI update strategy in slowly time-varying IoT channel scenarios, it is unnecessary for the BS to estimate the CSI of all the MTDs in every frame, which can reduce the training overhead significantly.

\begin{figure}[t]
\captionsetup{font={footnotesize, color = {black}}, justification = raggedright}
\centering
\includegraphics[scale=0.62]{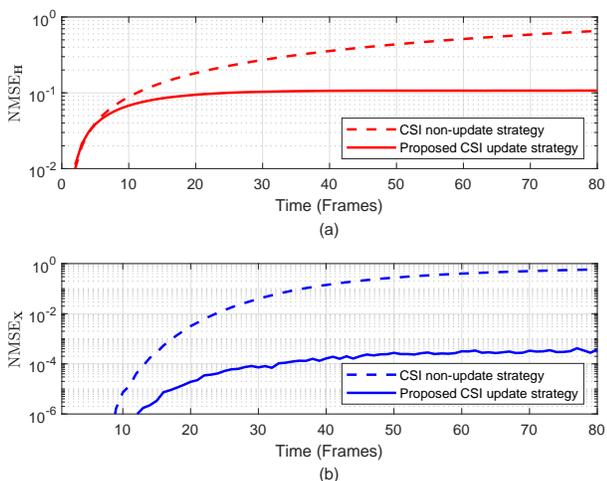}
\captionsetup{font={footnotesize, color = {black}}, justification = raggedright,labelsep=period}
\vspace{-2mm}
\caption*{Fig. 13. Performance comparison of the proposed CSI update strategy versus time (i.e., the number of frames) at SNR~$=$~30 dB and the AR coefficient $\alpha=0.99$: (a) ${\rm NMSE}_{\bf H}$ performance comparison; (b) ${\rm NMSE}_{\bf X}$ performance comparison.}
\vspace{-7mm}
\end{figure}

\vspace{-2mm}
\section{Conclusions}
This paper investigated the DADD problem for media modulation based mMTC, where both uncoded and coded transmissions were considered. By exploiting the {\it doubly structured sparsity} of media modulation signals, we first proposed a DS-AMP algorithm to solve the DADD problem for the uncoded case. Also, we derived the SE of the DS-AMP algorithm to theoretically predict its performance. Furthermore, for the coded case, we developed a BICMM scheme and proposed an IDS-AMP scheme based on SIC for improving the data decoding performance. In particular, a signature sequence part for facilitating the SIC processing was proposed to be embedded into the dedicated data packet and then the IDS-AMP detector was developed, whereby the estimated signal components were iteratively subtracted from the received signals for enhancing the data decoding performance. {\color{black}In addition, we discussed the CE problem and developed a data-aided CSI update strategy to reduce the training overhead in block fading channels.} Finally, extensive simulation results demonstrated the superiority of the proposed DS-AMP algorithm over cutting-edge algorithms in terms of ADER, SER, and BER performance while ensuring a lower computational complexity in the uncoded transmission. Improved data decoding performance of the proposed SIC-based IDS-AMP scheme was verified in the coded case as well. Also, simulation results verified the efficiency of the proposed CSI update strategy.

\vspace{-2mm}%
\appendix
{\color{black}Here we provide the detailed derivation of the EM update rules in the proposed algorithms.} According to (\ref{eq:EM1}) and (\ref{eq:EM2}), ${\rm ln}~p\left({{\bf X, Y};{\bm \theta}}\right)$ can be expressed as
\begin{equation}\label{eq:DerivationEM}
\begin{split}
&{\rm ln}~p\left({{\bf X, Y};{\bm \theta}}\right)\\
&={\rm ln}~p\left({{\bf Y}|{\bf X};\sigma_w^2}\right)+{\rm ln}~p\left({{\bf X; a}}\right)\\
&=\sum\nolimits_{j=1}^{J}\left[{{\rm ln}~p\left({{\bf y}_j|\tilde{\bf x}_j;\sigma_w^2}\right)+{\rm ln}~p\left({\tilde{\bf x}_j;{\bf a}}\right)}\right]\\
&=\sum\limits_{j=1}^{J}\left[{\sum\limits_{n=1}^{N_r}{\rm ln}~p\left({\left[{{\bf y}_j}\right]_n|\tilde{\bf x}_j;\sigma_w^2}\right)\!\!+\!\sum\limits_{k=1}^{K}{\rm ln}\!~p\left({{\bf x}_{k,j};a_k}\right)}\right].
\end{split}
\end{equation}

In order to find the parameter set $\bm \theta$ that maximizes $Q\left({{\bm \theta},{\bm \theta}^t}\right)$ in the $(t+1)$-th iteration, we differentiate $Q\left({{\bm \theta},{\bm \theta}^t}\right)$ with respect to $\bm \theta$ and let it equal zero.

First, according to (\ref{eq:PriorGeneral}) and (\ref{eq:DerivationEM}), we differentiate $Q\left({{\bm \theta},{\bm \theta}^t}\right)$ with respect to $a_k$, $\forall k$, as follows
\begin{equation}\label{eq:DerivationEMactivity1}
\begin{split}
&\dfrac{\partial Q\left({{\bm \theta},{\bm \theta}^t}\right)}{\partial a_k}=\dfrac{\partial \mathbb{E}\left\{{{\rm ln}~p\left({{\bf X, Y};{\bm \theta}}\right)|{\bf Y};{\bm \theta}^t}\right\}}{\partial a_k}\\
&=\sum\nolimits_{j=1}^{J}\mathbb{E}\left\{{\dfrac{d}{d a_k}{\rm ln}p\left({{\bf x}_{k,j};a_k}\right)|{\bf Y};{\bm \theta}^t}\right\}\\
&=\sum\nolimits_{j=1}^{J}\sum\nolimits_{{\bf x}_{k,j}\in \Gamma}p\left({{\bf x}_{k,j}|{\bf Y};{\bm \theta}^t}\right)\dfrac{d}{d a_k}{\rm ln}p\left({{\bf x}_{k,j};a_k}\right)\\
&\approx \sum\limits_{j=1}^{J}\sum\limits_{{\bf x}_{k,j}\in \Gamma}q\left({{\bf x}_{k,j}|r_{k,j}^t,\phi_{k,j}^t;\left({\sigma_w^2}\right)^t,a_k^t}\right)\dfrac{d}{d a_k}{\rm ln}p\left({{\bf x}_{k,j};a_k}\right)\\
&=\!\!\!\sum\limits_{j=1}^{J}\!\sum\limits_{{\bf x}_{k,j}\in \Gamma}\!\prod\limits_{i=1}^{N_t}\!\!q\left({\left[{{\bf x}_{k,j}}\right]_i|r_{k,j}^t,\!\phi_{k,j}^t;\!\left({\sigma_w^2}\right)^t,\!a_k^t}\right)\!\!\dfrac{d}{d a_k}{\rm ln}p\left({{\bf x}_{k,j};a_k}\right),
\end{split}
\end{equation}
where $\Gamma=\left\{{\Gamma_0,{\bf 0}_{N_t\times 1}}\right\}$ and $\Gamma_0$ is defined in (\ref{eq:MediaModulated_SignalSet}). The function $q\left({\left[{{\bf x}_{k,j}}\right]_i|r_{k,j}^t,\phi_{k,j}^t;\left({\sigma_w^2}\right)^t,a_k^t}\right)$ can be calculated according to (\ref{eq:PosMarginalAMP}) and
\begin{equation}\label{eq:DerivationEMactivity2}
\begin{array}{l}
\dfrac{d}{d a_k}{\rm ln}p\left({{\bf x}_{k,j};a_k}\right)=\dfrac{-\prod\nolimits_{i=1}^{N_t}\delta\left({\left[{{\bf x}_{k,j}}\right]_i}\right)}{p\left({{\bf x}_{k,j}; a_k}\right)}+\\
~~~~~~\dfrac{\dfrac{1}{N_t}\sum\nolimits_{i=1}^{N_t}\left[{\dfrac{1}{M}\sum\limits_{s\in\mathbb{S}}\delta\left({\left[{{\bf x}_{k,j}}\right]_i-s}\right)\prod\limits_{n\in [N_t],n\neq i}\delta\left({\left[{{\bf x}_{k,j}}\right]_n}\right)}\right]}{p\left({{\bf x}_{k,j};a_k}\right)}.
\end{array}
\end{equation}

Hence, the EM update rule for $a_k$, $\forall k$, can be expressed as
\begin{equation}\label{eq:EMactivityIndicaAppendix}
\begin{array}{l}
a_k^{t+1}=\dfrac{1}{J}\sum\limits_{j=1}^{J}\sum\limits_{{\bf x}_{k,j}\in \Gamma_0}\prod\limits_{i=1}^{N_t}q\left({\left[{{\bf x}_{k,j}}\right]_i|r_{l,j}^t, \phi_{l,j}^t;a_k^t}\right).
\end{array}
\end{equation}

Next, define ${\bf z}_j={\bf H}\tilde{\bf x}_j$ ($\forall j\in[J]$), then we have $\left[{{\bf z}_j}\right]_n=\sum\limits_{k=1}^{K}\left[{{\bf H}_k{\bf x}_{k,j}}\right]_n$ ($n\in N_r$). According to (\ref{eq:LikelihoodGeneral}) and (\ref{eq:DerivationEM}), we differentiate $Q\left({{\bm \theta},{\bm \theta}^t}\right)$ with respect to $\sigma_w^2$ as follows
\begin{equation}\label{eq:DerivationEMsigma1}
\begin{split}
&{\dfrac{\partial Q\left({{\bm \theta},{\bm \theta}^t}\right)}{\partial \left({\sigma_w^2}\right)}}=\dfrac{\partial \mathbb{E}\left\{{{\rm ln}~p\left({{\bf X, Y};{\bm \theta}}\right)|{\bf Y};{\bm \theta}^t}\right\}}{\partial \left({\sigma_w^2}\right)}\\
&=\sum\limits_{j=1}^{J}\sum\limits_{n=1}^{N_r}\mathbb{E}\left\{{\dfrac{d}{d \left({\sigma_w^2}\right)}{\rm ln} p\left({\left[{{\bf y}_j}\right]_n|\tilde{\bf x}_j;\sigma_w^2}\right)|{\bf Y};{\bm \theta}^t}\right\}\\
&=\sum\limits_{j=1}^{J}\sum\limits_{n=1}^{N_r}\mathbb{E}\left\{{ \dfrac{d}{d\left({\sigma_w^2}\right)}\left[{-{\rm ln}\left({\sigma_w^2}\right)-\dfrac{\left({\left[{{\bf y}_j}\right]_n-\left[{{\bf z}_j}\right]_n}\right)^2}{\left({\sigma_w^2}\right)}}\right]|{\bf Y};{\bm \theta}^t}\right\}.\\
\end{split}
\end{equation}

Hence, we have
\begin{equation}\label{eq:DerivationEMsigma2}
\begin{array}{l}
\left({{\sigma_w^2}}\right)^{t+1}=\dfrac{1}{JN_r}\sum\limits_{j=1}^{J}\sum\limits_{n=1}^{N_r}\mathbb{E}\left\{{\left({\left[{{\bf y}_j}\right]_n-\left[{{\bf z}_j}\right]_n}\right)^2|{\bf Y};{\bm \theta}^t}\right\},
\end{array}
\end{equation}
where for $n\in[N_r]$,
\begin{equation}\label{eq:DerivationEMsigma3}
\begin{array}{l}
\mathbb{E}\left\{{\left({\left[{{\bf y}_j}\right]_n-\left[{{\bf z}_j}\right]_n}\right)^2|{\bf Y};{\bm \theta}^t}\right\}\\
=\left({\left[{{\bf y}_j}\right]_n-\mathbb{E}\left\{{\left[{{\bf z}_j}\right]_n|{\bf y}_j;{\bm \theta}^t}\right\}}\right)^2
+{\rm Var}\left\{{\left[{{\bf z}_j}\right]_n|{\bf y}_j;{\bm \theta}^t}\right\},
\end{array}
\end{equation}
and ${\rm Var}\left\{{\cdot|{\bf y}_j;{\bm \theta}^t}\right\}$ is the variance conditioned on ${\bf y}_j$ with parameter ${\bm \theta}^t$.

Note that the distribution of $\left[{{\bf z}_j}\right]_n$ can be expressed as
\begin{equation}\label{eq:DerivationEMsigma4}
\begin{split}
&f\left({\left[{{\bf z}_j}\right]_n|{\bf y}_j;{\bm \theta}^t}\right)\\
&=\dfrac{1}{C}f\left({\left[{{\bf y}_j}\right]_n|\left[{{\bf z}_j}\right]_n;{\bm \theta}^t}\right)f\left({\left[{{\bf z}_j}\right]_n;{\bm \theta}^t}\right)\\
&=\dfrac{1}{C}{\cal CN}\left({\left[{{\bf y}_j}\right]_n;\left[{{\bf z}_j}\right]_n,\left({\sigma_w^2}\right)^t}\right){\cal CN}\left({\left[{{\bf z}_j}\right]_n;Z_{n,j}^t,V_{n,j}^t}\right)\\
&={\cal CN}\left({\left[{{\bf z}_j}\right]_n;\mathbb{E}\left\{{\left[{{\bf z}_j}\right]_n|{\bf y}_j;{\bm \theta}^t}\right\},{\rm Var}\left\{{\left[{{\bf z}_j}\right]_n|{\bf y}_j;{\bm \theta}^t}\right\}}\right),
\end{split}
\end{equation}
where
\begin{align}
\label{eq:DerivationEMsigma5} \mathbb{E}\left\{{\left[{{\bf z}_j}\right]_n|{\bf y}_j;{\bm \theta}^t}\right\}&=\dfrac{\left({\sigma_w^2}\right)^tZ_{n,j}^t+\left[{{\bf y}_j}\right]_nV_{n,j}^t}{\left({\sigma_w^2}\right)^t+V_{n,j}^t},\\
\label{eq:DerivationEMsigma6} {\rm Var}\left\{{\left[{{\bf z}_j}\right]_n|{\bf y}_j;{\bm \theta}^t}\right\}&=\dfrac{\left({\sigma_w^2}\right)^tV_{n,j}^t}{\left({\sigma_w^2}\right)^t+V_{n,j}^t},
\end{align}
and $C$ is the normalization factor, $V_{n,j}^t$ and $Z_{n,j}^t$ are calculated in (\ref{eq:UpdateV}) and (\ref{eq:UpdateZ}), respectively.

Hence, according to (\ref{eq:DerivationEMsigma2}), (\ref{eq:DerivationEMsigma3}), (\ref{eq:DerivationEMsigma5}), and (\ref{eq:DerivationEMsigma6}), we obtain the EM update rule of the noise variance as
\begin{equation}\label{eq:EMnoiseVarAppendix}
\begin{array}{l}
\left({\sigma_w^2}\right)^{t+1}=\dfrac{1}{JN_r}\sum\limits_{j=1}^{J}\sum\limits_{n=1}^{N_r}\left[{\dfrac{\left({\left[{{\bf y}_j}\right]_n-Z_{n,j}^t}\right)^2}{\left({1+\frac{V_{n,j}^t}{\left({\sigma_w^2}\right)^t}}\right)^2}+\dfrac{\left({\sigma_w^2}\right)^t V_{n,j}^t}{V_{n,j}^t+\left({\sigma_w^2}\right)^t}}\right].
\end{array}
\end{equation}

\end{document}